**Tribocharging of Granular Materials Dictated by Geometric and Thermoelectric Effects**

Donggeun Kang[1,†], Ji-Ho Mun[1], Seungil Baek[1], Jaeuk Seo[1], Sirichok Jungthawan[2], Yong Woon Kim[1], and Yong-Hyun Kim[1,2,*]

[1]*Department of Physics, Korea Advanced Institute of Science and Technology, Daejeon 34141, Republic of Korea*

[2]*School of Physics, Institute of Science, Suranaree University of Technology, Nakhon Ratchasima 30000, Thailand*

[†]Present address: Department of Physics and Chemistry, Korea Military Academy, Seoul, 01805, Republic of Korea

*Corresponding author: yong.hyun.kim@kaist.ac.kr

Tribocharging of granular materials is a widespread phenomenon ranging from volcanic eruptions and dust storms to pharmaceutical processing, yet its underlying mechanism remains elusive. Recent thermoelectric models have provided a quantitative framework for triboelectric charge transfer, but they have difficulty accounting for charging between identical materials. In this Letter, we propose a theory of granular tribocharging based on the interplay between particle geometry and thermoelectric response. Collision-induced frictional heating produces a size-dependent temperature asymmetry between particles, generating a thermoelectric voltage that drives charge transfer. The sign of the Seebeck coefficient determines the charging polarity, while particle size and thermal properties set its magnitude. Our multi-collision theory explains the experimentally observed size- and material-dependent reversal of charging polarity, as in quartz and nylon.

Tribocharging, the generation of electric charge through contact or friction between materials, is pervasive in both daily life and industrial systems. In granular materials, it plays a crucial role in natural phenomena such as volcanic eruptions, dust storms, and lightning [1], as well as in industrial processes, including electrophotography [2], spray coating [3], pharmaceutical manufacturing, and even coffee grinding [4,5]. Despite its ubiquity, the fundamental mechanism of granular tribocharging remains unresolved because it depends on multiple factors, including intrinsic material properties, particle morphology, and environmental conditions such as humidity and gas composition.

It has been commonly believed that granular insulators exhibit size-dependent bipolar charging, in which small particles charge negatively and large particles positively. Historically, this phenomenon was attributed by Lacks and co-workers to the transfer of high-energy electrons trapped at the surface [6–10]. However, experimental measurements by Waitukaitis *et al.* showed that the trapped-electron population was insufficient to account for the measured charge transfer, casting doubt on the universality of the trapped-electron hypothesis [11]. In the aftermath of this result, Lee *et al.* proposed an alternative carrier source, an ion transfer mediated by adsorbed $OH^-$ layers [12]. While this mechanism explains certain hydrophilic-hydrophobic contrasts, it cannot describe why chemically identical grains separate by polarity solely due to their size, nor why many polymers exhibit material-dependent polarity. Consequently, the microscopic origin of granular tribocharging has emerged as an open theoretical landscape since 2018, prompting a surge of new microscopic proposals—including mechanochemistry, strain-driven and flexoelectric polarization, tribo-tunneling—as well as recent evidence that adventitious carbonaceous adsorbates can act as hidden surface-state variables [13–18]. However, none of these frameworks quantitatively and simultaneously explain the two experimentally robust signatures that any unified theory must reproduce: (i) size-dependent polarity between identical particles [6–9,19] and (ii) material-dependent polarity reversal [11,20–22]. Although recent stochastic-scaling models reproduce variable impact charge, charge reversal, and size-dependent bipolar charging in large-scale particle simulations [23], they rely on experimentally prescribed impact-charge statistics rather than deriving the microscopic driving potential. Thus, the physical origin that unifies these macroscopic observations remains unresolved [15,24–32].

It would be natural to revisit the fundamental physical consequence of friction itself—heat generation at the interfaces—as a potential driver of tribocharging [33,34]. In particular, Zhang *et al.* demonstrated that frictional heating at a metal-semiconductor interface generates a steady thermoelectric voltage superimposed with a fluctuating tribovoltaic component [35]. Recent works on interfacial thermoelectricity indeed suggest that friction-induced temperature gradients produce a potential difference between surfaces. Shin *et al*. reported that the triboelectric factor $\xi = S/\sqrt{\rho c k}$— composed of material density $\rho$, specific heat $c$, thermal conductivity $k$, and Seebeck coefficient $S$— quantitatively accounts for the tribocharging behavior ranging from insulators to metals [13–15]. However, these thermoelectric models rely solely on intrinsic properties of semi-infinite materials in contact and exclude geometric effects [36–38], limiting their ability to explain the charging between chemically identical granular particles.

In this Letter, we reveal that granular tribocharging originates from geometry-dependent thermoelectric effects, providing a unified explanation for both size-dependent and material-dependent polarity reversal. By solving the heat diffusion equation during the collision process, the surface potential can then be evaluated from the resulting surface temperature profile as a function of the geometric and intrinsic factors: particle size $R$ and material properties ($\rho$, $c$, $k$, and $S$). Finally, we demonstrate that the experimentally observed size-dependent polarity trend in multiparticle collision is governed by these geometric and thermoelectric effects, identifying the dominant mechanism of granular tribocharging between identical materials.

When two particles collide, we assume that frictional heat is symmetrically injected into the two contacting surfaces [13,39]. This heat diffuses into the bulk, generating a transient temperature gradient. Because the extent of heat diffusion differs between the two particles, an asymmetric surface temperature profile develops, which in turn produces an electrostatic potential difference between the surfaces, as illustrated in Fig. 1(a). The key step in analyzing tribocharging tendencies within the thermoelectric framework is to determine the surface temperature difference $\Delta T$ that drives the potential difference $\Delta V = -S\Delta T$. This requires solving the heat diffusion equation given in Eq. (1), which describes how frictional heat diffuses from the contact region into the particle interior, imposing the Robin boundary condition (B.C.) to account for thermal interaction with the surrounding environment:

$$\frac{\partial T}{\partial t} = \frac{k}{\rho c}\nabla^2 T + \frac{\dot{q}}{\rho c} \tag{1}$$

$$(\text{B.C.})\ \ k\frac{dT}{dr}\Big|_{r=R} = -\kappa(T_{r=R} - T_{\infty}),$$

where $k$ is the intrinsic thermal conductivity and $\kappa$ is the effective interfacial conductance. We define the thermal barrier thickness as $\lambda \equiv k/\kappa$, which characterizes interfacial heat dissipation: a large $\lambda$ implies strong heat confinement in the material, whereas a small $\lambda$ corresponds to facile heat leakage to the environment. When the surface temperature asymmetry $\Delta T$ arises between colliding particles, the sign of $S$ determines two distinct cases, as illustrated in Figs. 1(b) and 1(c): (i) for $S > 0$, hole carriers transfer from the relatively cold particle to the hotter particle; (ii) for $S < 0$, the transfer direction is reversed compared to the positive-$S$ case by electron carriers. These demonstrate that the direction of charge transfer is determined by the sign of $S$. We assume that charge transfer adjusts instantaneously to the thermoelectric potential established by the local temperature difference during contact.

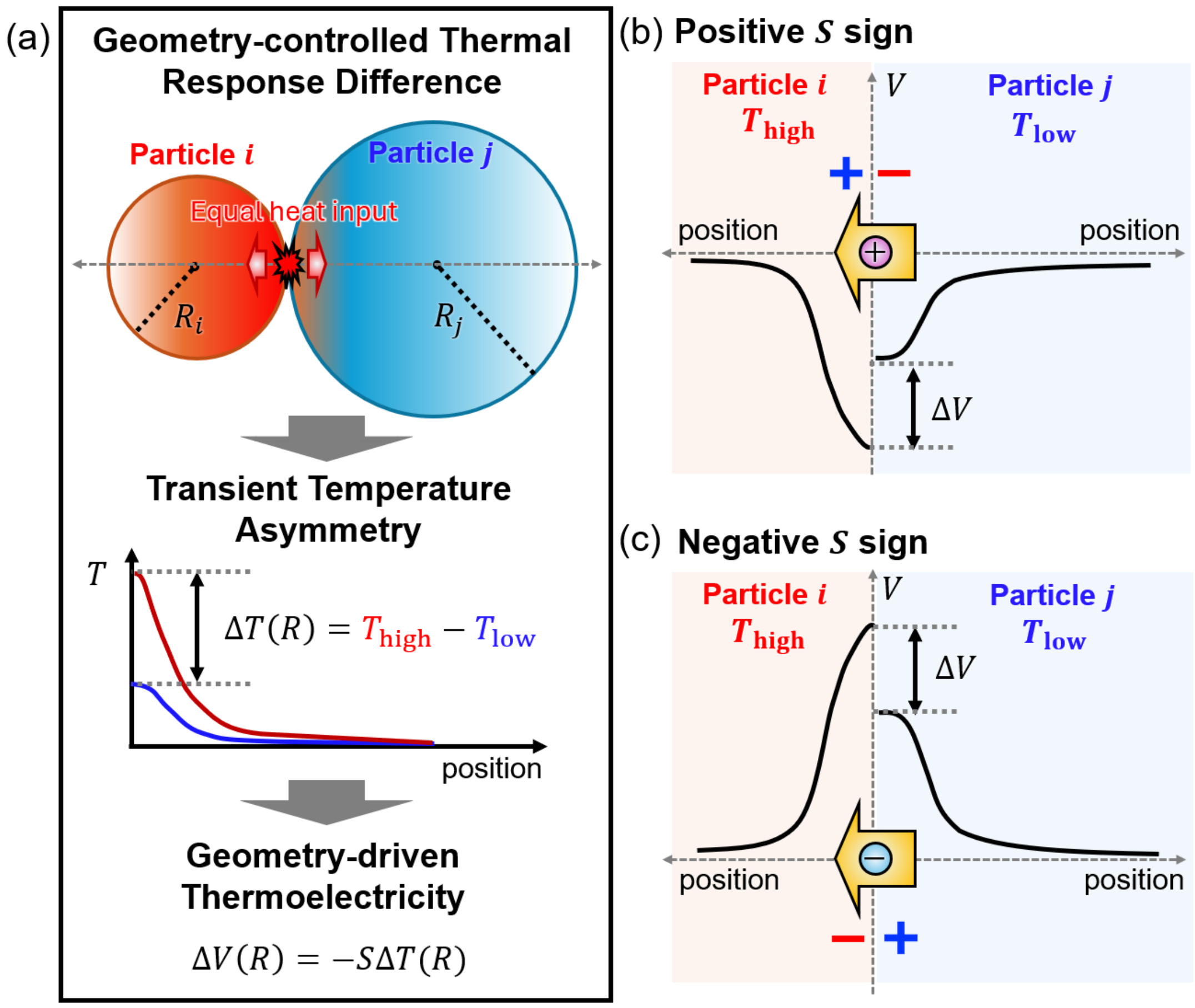

**Figure 1. Thermoelectric model for granular tribocharging.** (a) Upon collision, frictional heating with thermal response difference generates asymmetric surface temperature profiles through thermal diffusion, leading to a thermoelectric potential difference $\Delta V = -S\Delta T$. (b-c) Schematic potential profiles showing that the sign of Seebeck coefficient $S$ determines the direction of charge transfer: (b) positive and (c) negative sign. A positive (negative) Seebeck coefficient corresponds to hole-like (electron-like) charge carriers.

To model this process, we consider a simplified contact model in which colliding granular particles are assumed to be ideal spheres. Within the idealized Hertz contact theory [40–42], the contact geometry is determined by the radii $(R_i, R_j)$ of the colliding particle pair through the effective radius $R_{\mathrm{eff}}$, defined by $1/R_{\mathrm{eff}} = 1/R_i + 1/R_j$. The collision-induced heat input is modeled as a localized interfacial source with a Gaussian angular profile distributed over an effective contact area $A_{ij} \propto \pi R_{\mathrm{eff}}^2$. The symmetric heat-input assumption for the single collision calculations is implemented by assigning the heat-input energy density $Q_0 = 10^{-3}$ J cm$^{-2}$ for each contact surface. Accordingly, the total heat supplied to each particle during a collision is $Q_{ij} = Q_0 A_{ij}$, allowing the heat input to vary naturally with the pair geometry. The resulting heat diffusion problem is expanded in spatial eigenmodes $\phi_{nl}(r, \theta)$, where $n$ and $l$ label the radial and angular modes, respectively. The detailed eigenfunctions and modal derivation provided in the Supplemental Material [39].

Using the thermocouple relation $V = -ST$, the surface potential generated on particle $i$ during its collision with particle $j$ is then written as

$$V_i^{(ij)}(R_i, \theta, \tau_c) = -\frac{Q_{ij}}{\tau_c R_i}\frac{S}{k}\sum_{n,l} a_{nl}\phi_{nl}(R_i, \theta)\left(1 - \exp\left[-\frac{\tau_c}{\tau_{nl}}\right]\right), \tag{2}$$

where $\tau_c$ denotes the collision duration, and $Q_{ij}/\tau_c$ is the collision-induced heat input rate. The coefficient $a_{nl}$ represents the normalized projection of the localized heat source onto the eigenmode $\phi_{nl}$. The time evolution factor $1 - \exp[-\tau_c/\tau_{nl}]$ describes the fraction of the steady-state modal amplitude attained during the finite collision duration, where $\tau_{nl} = \rho c R^2/(k\mu_{nl}^2)$ is the relaxation time of $(n, l)$th mode. Here, $\mu_{nl}$ is the dimensionless eigenvalue with the $(n, l)$th thermal mode determined by the Robin boundary condition.

Equation (2) can be interpreted in terms of three key physical contributions. First, the prefactor $Q_{ij}/\tau_c R$ reflects the overall geometric scale, reflecting how strongly particle curvature controls the localization of injected heat at the surface. Smaller particles have larger curvature, which concentrates the injected heat and produces a stronger local thermal response. Second, the factor $S/k$ represents the intrinsic thermoelectric response, determining the magnitude and sign of the voltage, and thus the direction of charge transfer. Third, the normalized heat-source projection coefficient $a_{nl}$ specifies how strongly the localized heat source excites each thermal eigenmode. $a_{nl}$ is proportional to $[\mu_{nl}^2 + \mathrm{Bi}_i^2 - \mathrm{Bi}_i - l(l+1)]^{-1}$, where the Robin boundary condition enters $a_{nl}$ through the Biot number $\mathrm{Bi}_i = R_i/\lambda = \kappa R_i/k$, both explicitly and through the eigenvalue $\mu_{nl}$. In the limit of $\mathrm{Bi} \to 0$, the system approaches a thermally confining Neumann boundary condition, under which the temperature increases linearly with time. In contrast, increasing Bi strengthens heat exchange with the surroundings and shifts the modal response toward the heat-loss-dominated Dirichlet regime.

In essence, Eq. (2) shows that the surface potential is governed by the interplay between system size $R$ and intrinsic material properties ($\rho$, $c$, $k$, and $S$). Geometry enters not only through the explicit factor $1/R$, but also implicitly through the confinement ratio $R/\lambda$ and the relaxation time $\tau_{nl} \sim R^2$, thereby shaping the spatiotemporal evolution of the thermal response. Consequently, even in collisions between identical materials, a finite charge transfer can arise purely from geometry-induced surface temperature asymmetry, establishing a *geometry-driven thermoelectric* description of granular tribocharging.

To disentangle the intrinsic contributions from geometric effects, we first consider particle pairs of identical size composed of different materials. This allows us to isolate the material-dependent thermodynamic effects from the overall thermal response under otherwise identical collision conditions. For all materials considered in this study, the values of $\rho$, $c$, $k$, and $S$ are taken from Ref. [13]. In Fig. 2(a), the surface temperature profiles for a collision between two particles with the same radius ($R_i = 100$ μm) evaluated at a fixed collision time ($\tau_c = 10^{-4}$ s) [43] exhibit clear material-dependent variations. This demonstrates that intrinsic thermal properties ($\rho$, $c$, and $k$) alone can give rise to a finite temperature difference within the frictional contact region, generalizing the previous thermoelectric tribocharging model for semi-infinite planes [14]. Figure 2(b) illustrates that the surface potential $V$ shows pronounced variations in both magnitude and polarity across materials, indicating that the sign and magnitude of $S$ play a key role in determining the polarity of the generated voltage. To illustrate this material contrast, we select nylon and quartz—widely used in tribocharging experiments [6–9,19,44]—as representative polymeric and mineral materials with opposite signs of $S$ [Fig. 2(c)]. In the absence of geometric asymmetry, the *thermoelectric factor* $S/k$ governs the collision-induced thermoelectric response between dissimilar particles. Accordingly, the triboelectric series can be controlled by $S/k$ [Fig. 2(d)]. This sequence indeed follows a similar trend to the triboelectric factor $\xi = S/\sqrt{\rho c k}$ reported in previous studies, and the predicted polarity ordering agrees with 83% of the reported triboelectric series [13].

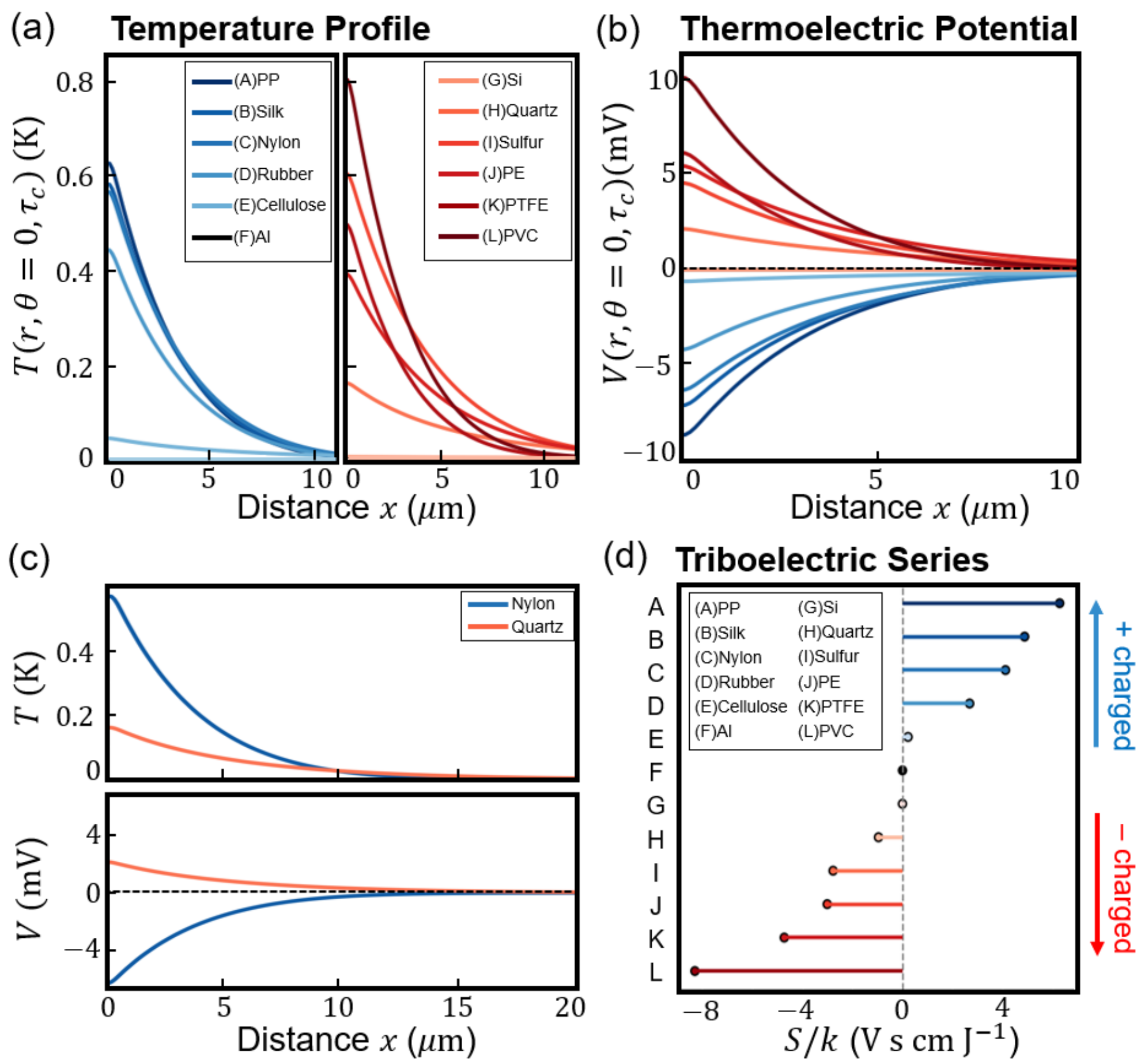

**Figure 2. Material-dependent thermoelectric response under identical collision geometry.** (a) Temperature profile $T(r, \theta; \tau_c)$ at $\tau_c = 10^{-4}$ s for equal-sized particles ($R_i = R_j = 100$ μm), where $x$ denotes the distance from the contact surface. Despite identical geometry and heat input, different materials exhibit distinct temperature distributions, corresponding to positive and negative $S$, respectively. (b) The resulting potential $V(r, \theta; \tau_c)$ shows pronounced material-dependent variations in both magnitude and polarity. (c) Temperature and potential profiles for Nylon ($S > 0$) and Quartz ($S < 0$) (d) The thermoelectric response is governed by $S/k$, whose sign and magnitude determine the direction and strength of charge transfer.

Based on the contribution from intrinsic material properties, we examine the role of geometry by varying particle size. To explicitly demonstrate size-dependent behavior, we computed the surface temperature profiles for quartz pairs with different particle sizes. As shown in Fig. 3, the surface temperature difference increases as the size disparity between the two particles becomes larger. As discussed in Eq. (2), this behavior originates from geometric scale ($Q_{ij}/\tau_c R$) and thermal confinement (Bi). Smaller particles produce a stronger local temperature response through both the increased geometric factor and suppressed heat loss to the environment. Consequently, a greater size disparity enhances the temperature asymmetry between the colliding particles, inducing a larger thermoelectric voltage even between identical materials.

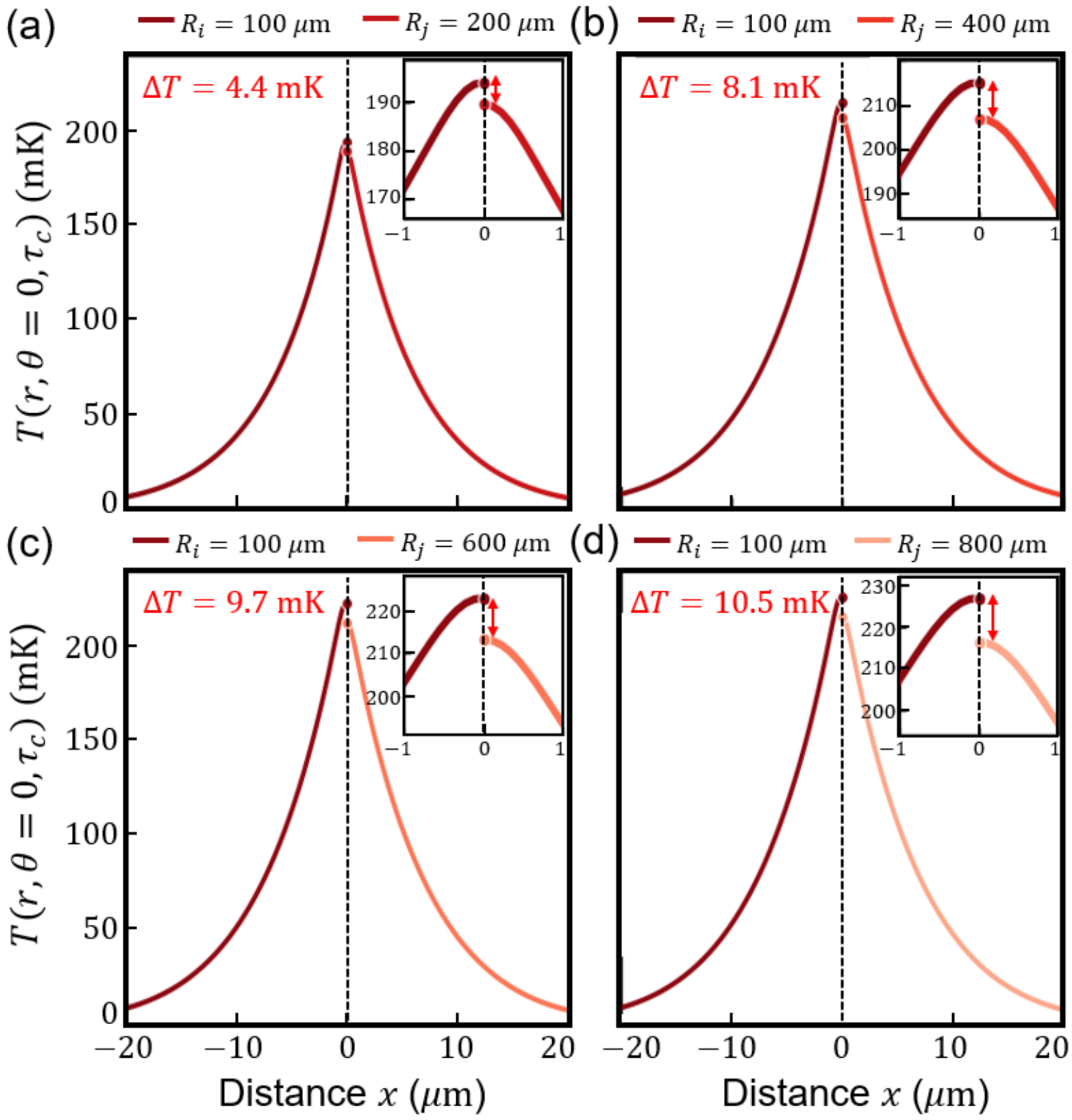

**Figure 3. Geometry-driven thermoelectric asymmetry in quartz.** (a-d) Temperature profiles $T(r, \theta = 0; \tau_c)$ for quartz particles with increasing size disparity $R_i = 100\ \mu m$, $R_j = 200$ - $800\ \mu m$, and $\tau_c = 10^{-4}$ s. The signed distance from the contact is defined as $x = r - R_i < 0$ within particle $i$ and $x = R_j - r > 0$ within particle $j$, with $x = 0$ at the contacting surfaces. The resulting contact-point temperature difference, $\Delta T = |T(R_i, 0; \tau_c) - T(R_j, 0; \tau_c)|$, increases with the radius ratio, reaching 10.5 mK at $R_j/R_i = 8$.

While Eq. (2) captures the thermal response during a single collision, granular tribocharging is inherently a repeated multi-collision process. Therefore, the relevant question is how this geometry-induced thermoelectric voltage accumulates over successive short collision events. For a representative particle radius of $R_i \sim 100\ \mu m$, the thermal conductivities of quartz ($1.4\ Wm^{-1}K^{-1}$) and nylon ($0.27\ Wm^{-1}K^{-1}$) [13], together with the prescribed interfacial conductance $\kappa = 1\ Wm^{-2}K^{-1}$, give Biot number of approximately $7.1 \times 10^{-5}$ and $3.7 \times 10^{-4}$, respectively. These small values place the present model in the Neumann regime. Assuming the collision duration $\tau_c$ is much shorter than the average inter-collision time, we can consider the particle-averaged response of the spatial eigenmode $\phi_{nl}$ as the accumulated effect of many collisions. Random collision orientations suppress the anisotropic angular contributions ($l \neq 0$), and the volume integral of thermal response almost averages out the higher radial modes ($n \neq 0$). The remaining lowest eigenmode ($n = l = 0$) is nearly uniform. Using $e^x \approx 1 + x$ for $\tau_c \ll \tau_{00}$, Eq. (2) effectively reduces to

$$\bar{V}_i^{(ij)} \simeq -\frac{Q_{ij}}{R_i^3}\frac{S}{\rho c}\mu_{00}^2 a_{00}\langle\phi_{00}\rangle_V \tag{3}$$

where $\langle\phi_{00}\rangle_V$ denotes the volume-averaged lowest spatial eigenmode. In this limit, $\mu_{00}^2 a_{00}\langle\phi_{00}\rangle_V$ approaches $3/(4\pi)$ (see Supplemental Material S3.A [39]). Eq. (3) therefore shows that each collision contributes to the multi-collision response through the thermal-inertia prefactor $1/(\rho c R^3)$, rather than an explicit thermal-conductivity $1/k$.

We construct a statistical accumulation model based on successive random binary collisions. The number of collisions between particles $i$ and $j$, $N_{ij}$, is sampled from a Poisson distribution with mean $N_{\text{tot}}P_{ij}$, where $N_{\text{tot}}$ is an expected total number of collision events and $P_{ij}$ is the normalized pair-collision probability proportional to $\left(R_i + R_j\right)^2$. To account for cumulative thermal loss without tracking individual collision histories, we introduce the effective thermal contribution per collision, $Q_{ij}^{(\text{mc})} = Q_0^{(\text{mc})}A_{ij}$, with $Q_0^{(\text{mc})} = 10^{-5}\ \text{J cm}^{-2}$. This phenomenological parameter also incorporates unresolved differences between local contact-temperature and particle-averaged responses, without assuming their equivalence. The cumulative pair-specific potential is then

$$V_i^{(ij)} = -\frac{3N_{ij}Q_{ij}^{(\text{mc})}}{4\pi R_i^3}\frac{S}{\rho c}. \tag{4}$$

Equation (4) shows that the same pair-specific effective thermal contribution generates different potentials owing to the different particle heat capacities. The resulting potential difference scales with $R_i^{-3} - R_j^{-3}$, with its magnitude governed by the collision statistics and contact geometry, and its polarity determined by the sign of the Seebeck coefficient $S$. The subsequent charge-transfer calculation is detailed in Supplemental Material S3.B [39].

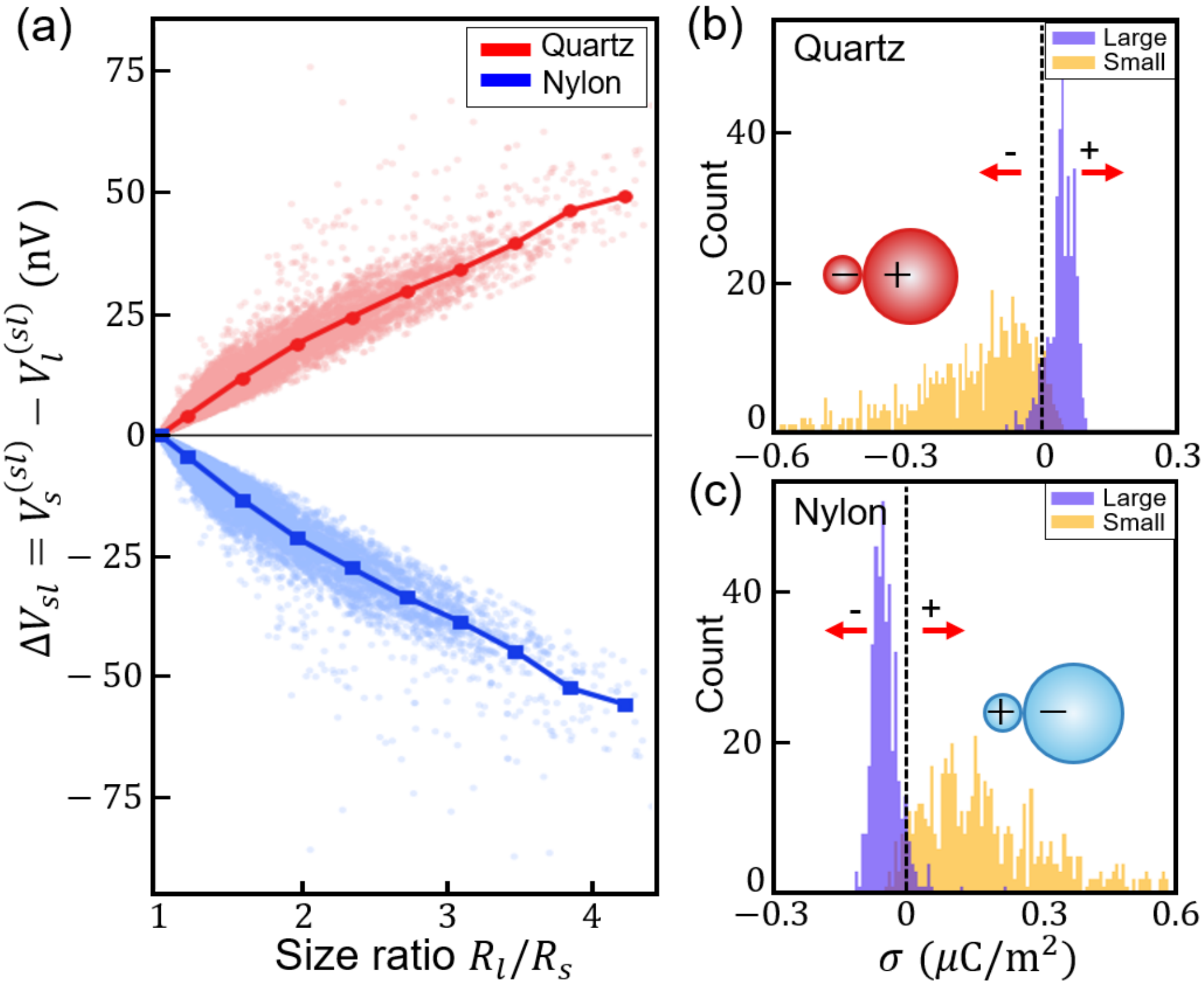

**Figure 4. Size-dependent pairwise voltage and charge segregation for Quartz and Nylon with opposite *S*** (quartz: red, nylon: blue). (a) Pairwise voltage $\Delta V_{sl} = V_s^{(sl)} - V_l^{(sl)} = -\frac{3N_{sl}Q_{sl}^{(\mathrm{mc})}}{4\pi}\frac{S}{\rho c}(R_s^{-3} - R_l^{-3})$, calculated from Eq. (4) for each active collision pair and plotted against the size ratio $R_l/R_s$, where $R_s = \min(R_i, R_j)$ and $R_l = \max(R_i, R_j)$. Solid lines with symbols show the corresponding average values. (b,c) Surface charge density $\sigma$ distributions obtained by converting these pairwise voltages into charge transfer contributions and accumulating them over $10^4$ collision events for quartz and nylon, respectively. Small and large particles acquire opposite polarities, with a reversal between materials, explaining the polarity reversal trend.

To investigate how this size-dependent driving force develops into macroscopic charge segregation, we next examine the charging behavior of quartz and nylon in a bimodal granular mixture consisting of 600 small ($50 \le R_s \le 150$ μm) and 400 large ($100 \le R_l \le 400$ μm) particles. The radii were sampled from a log-normal distribution with $\sigma_{\ln R} = 0.2$ to mimic realistic powders [45,46], and the expected total number of collision events was set to $N_{\mathrm{tot}} = 10^4$. For the pairwise analysis, we relabel particles $i$ and $j$ as $s$ and $l$ to explicitly denote the small and large particles of each pair, respectively. Figure 4(a) shows the resulting pairwise voltage $\Delta V_{sl} = V_s^{(sl)} - V_l^{(sl)}$ obtained by Eq. (4). Its magnitude increases on average with the size ratio $R_l/R_s$, while its spread at a given ratio reflects differences in the absolute particle sizes and collision counts. The corresponding pairwise charge transfers are then accumulated over all collision partners to obtain the surface charge density distributions. Figure 4(b) reproduces the conventional polarity trend for quartz, where the negative Seebeck coefficient leads small particles to charge negatively and large particles to charge positively [6–9]. For nylon, the positive Seebeck coefficient reverses the polarity, causing large particles to charge negatively, as shown in Fig. 4(c) [11,20–22,37,47]. These results confirm that the geometric effects govern size-dependent charge segregation, while material-dependent thermoelectric properties dictate the overall charging polarity.

In conclusion, we establish a unified framework for granular tribocharging based on the interplay between geometry and intrinsic thermoelectric properties. Upon particle collision, geometry-induced thermal asymmetry generates a thermoelectric voltage that explains the tribocharging between identical materials. A statistical study on multi-collision explains the long-standing issue of polarity reversal through the bipolar nature of the Seebeck coefficient. This study highlights thermoelectricity as a key mechanism in granular triboelectric charging of identical materials and offers a general strategy for controlling contact-induced charge transfer.

**ACKNOWLEDGEMENTS**

This work was supported by the Grand Challenge 30 program from the College of Natural Sciences, KAIST and by the National Research Foundation of Korea (NRF) grant funded by the Korea government (MSIT) (RS-2026-25595872).

**References**

[1] E. R. Williams, *The Electrification of Thunderstorms*, Sci. Am. **259**, 88 (1988).

[2] D. M. Pai and B.E. Springett, *Physics of electrophotography*, Rev. Mod. Phys. **65**, 163 (1993).

[3] A. G. Bailey, *The science and technology of electrostatic powder spraying, transport and coating*, J. Electrost. **45**, 85 (1998).

[4] K. C. Pingali, S. V. Hammond, F. J. Muzzio, and T. Shinbrot, *Use of a static eliminator to improve powder flow*, Int. J. Pharm. **369**, 2 (2009).

[5] J. M. Harper, C. S. McDonald, E. J. Rheingold, Y.-H. Kim, Josef Dufek and Christopher H. Hendon, *Moisture-controlled triboelectrification during coffee grinding*, Matter **7**, 266 (2024).

[6] D. J. Lacks and A. Levandovsky, *Effect of particle size distribution on the polarity of triboelectric charging in granular insulator systems*, J. Electrost. **65**, 107 (2007).

[7] K. M. Forward, D. J. Lacks, and R. M. Sankaran, *Charge Segregation Depends on Particle Size in Triboelectrically Charged Granular Materials*, Phys. Rev. Lett. **102**, 028001 (2009).

[8] K. M. Forward, D. J. Lacks, and R. M. Sankaran, *Triboelectric charging of granular insulator mixtures due solely to particle - Particle interactions*, Ind. Eng. Chem. Res. **48**, 2309 (2009).

[9] M. A. Bilici, J. R. Toth, R. M. Sankaran, and D. J. Lacks, *Particle size effects in particle-particle triboelectric charging studied with an integrated fluidized bed and electrostatic separator system*, Rev. Sci. Instrum. **85**, 103903 (2014).

[10] D. J. Lacks, N. Duff, and S. K. Kumar, *Nonequilibrium Accumulation of Surface Species and Triboelectric Charging in Single Component Particulate Systems*, Phys. Rev. Lett. **100**, 188305 (2008).

[11] S. R. Waitukaitis, V. Lee, J. M. Pierson, S. L. Forman, and H. M. Jaeger, *Size-Dependent Same-Material Tribocharging in Insulating Grains*, Phys. Rev. Lett. **112**, 218001 (2014).

[12] V. Lee, N. M. James, S. R. Waitukaitis, and H. M. Jaeger, *Collisional charging of individual submillimeter particles: Using ultrasonic levitation to initiate and track charge transfer*, Phys. Rev. Mater. **2**, 035602 (2018).

[13] E.-C. Shin, J.-H. Ko, H.-K. Lyeo, and Y.-H. Kim, *Derivation of a governing rule in triboelectric charging and series from thermoelectricity*, Phys. Rev. Research **4**, 023131 (2022).

[14] J.-H. Mun, E.-C. Shin, J. Seo, and Y.-H. Kim, *Triboelectric charge transfer theory driven by thermoelectric effect*, Phys. Rev. Research **7**, L032067 (2025).

[15] E.-C. Shin, J.-H. Mun, S. Baek, J. Jung, and Y.-H. Kim, *Recent progress in understanding the microscopic origin of triboelectricity from mechanochemistry to thermoelectricity*, Chem. Phys. Rev. **4**, 031305 (2023).

[16] J. C. A. Sobarzo, F. Pertl, D. M. Balazs, T. Costanzo, M. Sauer, A. Foelske, M. Ostermann, C. M. Pichler, Y. Wang, Y. Nagata, M. Bonn, and S. Waitukaitis, *Spontaneous ordering of identical materials into a triboelectric series*, Nature **638**, 664 (2025).

[17] J. Wu, X. Wang, H. Li, F. Wang, W. Yang, and Y. Hu, *Insights into the mechanism of metal-polymer contact electrification for triboelectric nanogenerator via first-principles investigations*, Nano Energy **48**, 607 (2018).

[18] G. Grosjean, M. Ostermann, M. Sauer, M. Hahn, C. M. Pichler, F. Fahrnberger, F. Pertl, D. M. Balazs, M. N. Link, S. H. Kim, D. L. Schrader, A. Blanco, F. Gracia, N. Mujica, and S. R.

Waitukaitis, *Adventitious carbon breaks symmetry in oxide contact electrification*, Nature **651**, 626 (2026).
[19] J. R. Toth, A. K. Phillips, S. Rajupet, R. M. Sankaran, and D. J. Lacks, *Particle-Size-Dependent Triboelectric Charging in Single-Component Granular Materials: Role of Humidity*, Ind. Eng. Chem. Res. **56**, 9839 (2017).
[20] J. Lowell and W. S. Truscott, *Triboelectrification of identical insulators. I. An experimental investigation*, J. Phys. D Appl. Phys. **19**, 1273 (1986).
[21] C. Han, Q. Zhou, J. Hu, C. Liang, X. Chen, and J. Ma, *The charging characteristics of particle–particle contact*, J. Electrost. **112**, 103582 (2021).
[22] D. Carter and C. Hartzell, *Effect of mixture properties on size-dependent charging of same-material dielectric grains*, J. Electrost. **107**, 103475 (2020).
[23] H. Grosshans, G. Ozler, V. Veeravalli, and S. Jantač, *Unifying same- and different-material particle charging through stochastic scaling*, Phys. Rev. X **16**, 011023 (2026).
[24] D. J. Lacks and T. Shinbrot, *Long-standing and unresolved issues in triboelectric charging*, Nat. Rev. Chem. **3**, 465 (2019).
[25] Y. Zhao, M. Liu, C. H. Wang, S. Matsusaka, and J. Yao, *Electrostatics of granules and granular flows: A review*, Adv. Powder Technol. **34**, 103895 (2023).
[26] S. Pan and Z. Zhang, *Fundamental theories and basic principles of triboelectric effect: A review*, Friction **7**, 2 (2019).
[27] D. M. Gooding and G. K. Kaufman, *Tribocharging and the Triboelectric Series*, in Encyclopedia of Inorganic and Bioinorganic Chemistry (John Wiley & Sons, Ltd., Hoboken, NJ, 2019), pp. 1–14.
[28] W. R. Harper, *Electrification following the contact of solids*, Contemp. Phys. **2**, 345 (1961).
[29] H. Zou et al., *Quantifying the triboelectric series*, Nat. Commun. **10**, 1427 (2020).
[30] H. Zou, L. Guo, H. Xue, Y. Zhang, X. Shen, X. Liu, P. Wang, X. He, G. Dai, P. Jiang, H. Zheng, B. Zhang, C. Xu, and Z. L. Wang, *Quantifying and understanding the triboelectric series of inorganic non-metallic materials*, Nat. Commun. **11**, 2093 (2020).
[31] X. Liu, J. Kolehmainen, I. Nwogbaga, A. Ozel, and S. Sundaresan, *Effect of particle size on tribocharging*, Powder Technol. **375**, 199 (2020).
[32] S. Matsusaka, H. Maruyama, T. Matsuyama, and M. Ghadiri, *Triboelectric charging of powders: A review*, Chem. Eng. Sci. **65**, 5781 (2010).
[33] P S H Henry, *The role of asymmetric rubbing in the generation of static electricity*, Br. J. Appl. Phys. **4**, S31 (1953).
[34] J. Lowell and W. S. Truscott, *Triboelectrification of identical insulators. II. Theory and further experiments*, J. Phys. D: Appl. Phys. **19**, 1281 (1986).
[35] Z. Zhang, T. He, J. Zhao, G. Liu, Z. L. Wang, and C. Zhang, *Tribo-thermoelectric and tribovoltaic coupling effect at metal-semiconductor interface*, Mater. Today Phys. **16**, 100295 (2021).
[36] S. Ge, Z. Fan, C. Fang, Z. Huang, J. Sun, J. Wang, Y. Yang, and Y. Yang, *Contact electrification polarity of polyethylene particles: A morphology-derived anomalous size dependence*, AIChE J. **69**, e18196 (2023).
[37] O. Verners, L. Lapčinskis, L. Ģermane, A. Kasikov, M. Timusk, K. Pudzs, A. V. Ellis, P. C. Sherrell, and A. Šutka, *Smooth polymers charge negatively: Controlling contact electrification polarity in polymers*, Nano Energy **104**, 107914 (2022).

[38] C. Xu, B. Zhang, A. C. Wang, H. Zou, G. Liu, W. Ding, C. Wu, M. Ma, P. Feng, Z. Lin, and Z. L. Wang, *Contact-electrification between two identical materials: Curvature effect*, ACS Nano **13**, 2034 (2019).
[39] See the Supplemental Material for details
[40] V. L. Popov, M. Heß, and E. Willert, *Handbook of Contact Mechanics* (Springer, Berlin, 2018).
[41] V. L. Popov, *Contact Mechanics and Friction: Physical Principles and Applications* (Springer, Berlin, 2010).
[42] G. C. McLaskey and S. D. Glaser, *Hertzian impact: Experimental study of the force pulse and resulting stress waves*, J. Acoust. Soc. Am. **128**, 1087 (2010).
[43] D. L. Blair and A. Kudrolli, *Collision statistics of driven granular materials*, Phys. Rev. E **67**, 041301 (2003).
[44] E.-S. Lee, S. Cho, H.-K. Lyeo, and Y.-H. Kim, *Seebeck Effect at the Atomic Scale*, Phys. Rev. Lett. **112**, 136601 (2014).
[45] P. Mehrani, M. Murtomaa, and D. J. Lacks, *An overview of advances in understanding electrostatic charge buildup in gas-solid fluidized beds*, J. Electrost. **87**, 64 (2017).
[46] P. S. Iliev, F. K. Wittel, and H. J. Herrmann, *Evolution of fragment size distributions from the crushing of granular materials*, Phys. Rev. E **99**, 012904 (2019).
[47] S. B. Rice, C. Chan, S. C. Brown, P. Eschbach, L. Han, D. S. Ensor, A. B. Stefaniak, J. Bonevich, A. E. Vladár, A. R. Hight Walker, J. Zheng, C. Starnes, A. Stromberg, J. Ye, and E. A. Grulke, *Particle size distributions by transmission electron microscopy: An interlaboratory comparison case study*, Metrologia **50**, 663 (2013).
[48] Y. Zhang, A. Ozel, and C. M. Hartzell, *A macroscopic donor–acceptor-based discrete element model for contact electrification of insulating granular materials*, Powder Technol. **431**, 119029 (2024).

# Supplemental Material:

# Tribocharging of Granular Materials Dictated by Geometric and Thermoelectric Effects

Donggeun Kang[1,†], Ji-Ho Mun[1], Seungil Baek[1], Jaeuk Seo[1], Sirichok Jungthawan[2], Yong Woon Kim[1], and Yong-Hyun Kim[1,2,*]

[1]*Department of Physics, Korea Advanced Institute of Science and Technology, Daejeon 34141, Republic of Korea*

[2]*School of Physics, Institute of Science, Suranaree University of Technology, Nakhon Ratchasima 30000, Thailand*

[†]Present address: Department of Physics and Chemistry, Korea Military Academy, Seoul, 01805, Republic of Korea

*Corresponding author: yong.hyun.kim@kaist.ac.kr

## S1. Single collision model and the temperature profile of finite collision model

### A. Hertzian contact theory and heat source model

According to the Hertzian contact theory [R1,R2], the contact radius $a_{ij}$ and mutual indentation depth $d_{ij}$ of two colliding elastic spheres are related by $a_{ij}^2 = R_{\text{eff}} d_{ij}$, where $R_{\text{eff}}$ is the effective radius of two colliding particles ($R_i$, $R_j$), defined as $1/R_{\text{eff}} = 1/R_i + 1/R_j$ (Fig. S1). For an elastic collision, $d_{ij}$ is generally determined by the effective mass, impact velocity, and effective Young's modulus. Rather than evaluating these collision-dynamic contributions separately for every particle-size ratio, we introduce the normalized indentation parameter $\eta = d_{ij}/R_{\text{eff}}$, and prescribe the same representative value, $\eta = 10^{-2}$, for all collision pairs. The contact geometry is therefore parameterized as

$$a_{ij} = \sqrt{R_{\text{eff}} d_{ij}}, \qquad d_{ij} = \eta R_{\text{eff}}, \qquad A_{ij} = \eta \pi R_{\text{eff}}^2. \tag{S1}$$

A fixed $\eta$ corresponds to imposing the same normalized indentation, $d_{ij}/R_{\text{eff}}$, across the particle pairs. This controlled parametrization allows the size-dependent thermal response to be examined independently of additional variations in velocity, elastic properties, and effective collision mass. Here, $\eta$ should be interpreted as a prescribed effective contact parameter for inelastic collisions, compared to a universal constant predicted by Hertzian elastic collision dynamics. Additionally, the collision duration is $\tau_c \sim 10^{-4}$ s, reflecting that collision events are much shorter than the experimentally measured inter-collision time scale (ms) [R3].

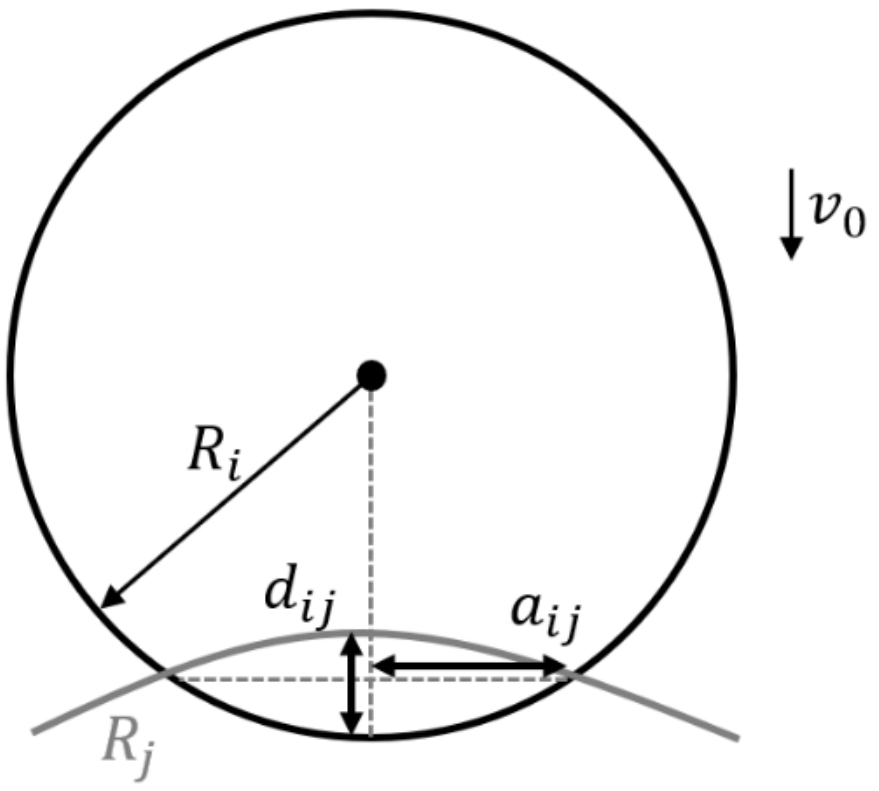

**Figure S1.** Schematic of the Hertzian contact geometry used to model a normal collision between two spherical particles $i$ and $j$, with radii $R_i$ and $R_j$, respectively. Their effective radius is defined as $1/R_{\text{eff}} = 1/R_i + 1/R_j$. Upon collision with relative normal velocity $v_0$, the particles undergo a mutual indentation $d_{ij}$ and form a circular contact region of radius $a_{ij}$. These geometric quantities satisfy the Hertzian relation $a_{ij}^2 = R_{\text{eff}} d_{ij}$. To compare the size-dependent thermal responses under a common dimensionless contact condition, we prescribe a fixed indentation ratio $\eta \equiv d_{ij}/R_{\text{eff}} = 10^{-2}$, yielding $d_{ij} = \eta R_{\text{eff}}$ and $A_{ij} = \pi a_{ij}^2 = \eta \pi R_{\text{eff}}^2$.

With these contact assumptions, we next construct the heat source generated during a single inelastic collision. In general, the instantaneous heat-generation rate may vary over the collision because the contact force and geometry evolve with time. Rather than resolving this detailed collision dynamics, we approximate the temporal heat input by a constant heat-input over the contact duration that preserves the total heat $Q_{ij}$ deposited over the contact duration $\tau_c$. Here, the total heat $Q_{ij}$ means the heat deposited into each particle of the colliding pair; hence the total interfacial heat assigned to the two-particle system is $2Q_{ij}$. Accordingly, the total heat-input rate of a single particle is taken to be $Q_{ij}/\tau_c$ during contact and zero after separation:

$$\dot{q}(r', \theta'; t) = \begin{cases} \dot{q}(r', \theta'), & 0 < t \leq \tau_c, \\ 0, & t > \tau_c. \end{cases} \tag{S2}$$

The spatial distribution of the heat input is modeled by separable Gaussian functions in the radial coordinate $r'$ and polar angle $\theta'$. The volumetric heat-input rate is therefore written as

$$\dot{q}(r',\theta') = \frac{Q_{ij}}{\tau_c \Lambda_{ij}} \exp\left(-\frac{(R_i - r')^2}{2\beta_{ij}^2}\right) \exp\left(-\frac{\theta'^2}{2\alpha_{ij}^2}\right) \tag{S3}$$

with the radial-angular normalization

$$\Lambda_{ij} = 2\pi \int_0^R \exp\left(-\frac{(R_i - r')^2}{2\beta_{ij}^2}\right) r'^2 dr' \int_0^\pi \exp\left(-\frac{\theta'^2}{2\alpha_{ij}^2}\right) \sin\theta' \, d\theta' . \tag{S4}$$

This normalization ensures that the spatially integrated heat-input rate during contact is $\int_V \dot{q}(r',\theta')dV' = Q_{ij}/\tau_c$ . Consequently, integration over the entire contact duration gives $\int_0^{\tau_c} \int_V \dot{q}(r',\theta';t)dV'\,dt = Q_{ij}$. The angular width $\alpha_{ij}$ is determined by mapping the contact area onto a spherical cap on particle $i$, yielding $A_{ij} = 2\pi R_i^2(1 - \cos\alpha_{ij})$, and thus $\alpha_{ij} = \cos^{-1}\left[1 - \frac{A_{ij}}{2\pi R_i^2}\right]$. The radial width $\beta_{ij} = d_{ij}/2$. Thus, $\alpha_{ij}$ and $\beta_{ij}$ specify the polar-angular and the radial widths of the localized heat source (Fig. S2), and they depend on the geometry of the collision pair through $R_{\text{eff}}$.

Furthermore, we define $Q_0 = 10^{-3}$ J cm$^{-2}$ as the areal heat-input energy density. The total heat supplied during a collision between particle $i$ and $j$ is then $Q_{ij} = Q_0 A_{ij}$. Accordingly, $Q_{ij}/\tau_c$ represents the total heat input rate delivered through the contact area $A_{ij}$ over the collision duration $\tau_c$. The magnitude of $Q_{ij}$ can be estimated from the collision energy scale. For a representative equal-sized pair with $R_i = R_j = 100\,\mu$m and $\eta = 10^{-2}$, the prescribed contact area gives $Q_{ij} \cong 7.9 \times 10^{-10}$J. Compared with the estimated collision energy scale of $10^{-8} - 10^{-7}$J — density $\rho \sim 1$–2 g/cm$^3$, and impact velocity $v_0 \sim 1$–5 m/s [R4] —this corresponds to a pair- and velocity-dependent conversion fraction of approximately 0.8–8%. This comparison provides an order-of-magnitude estimate of the prescribed heat input relative to the collision kinetic-energy scale, rather than a measured or microscopically determined heat-conversion efficiency. Over $\tau_c = 10^{-4}$s, the corresponding heat-input rate is $Q_{ij}/\tau_c \cong 10^{-6} - 10^{-5}$ J/s. This range defines the physically relevant scale used throughout the analysis.

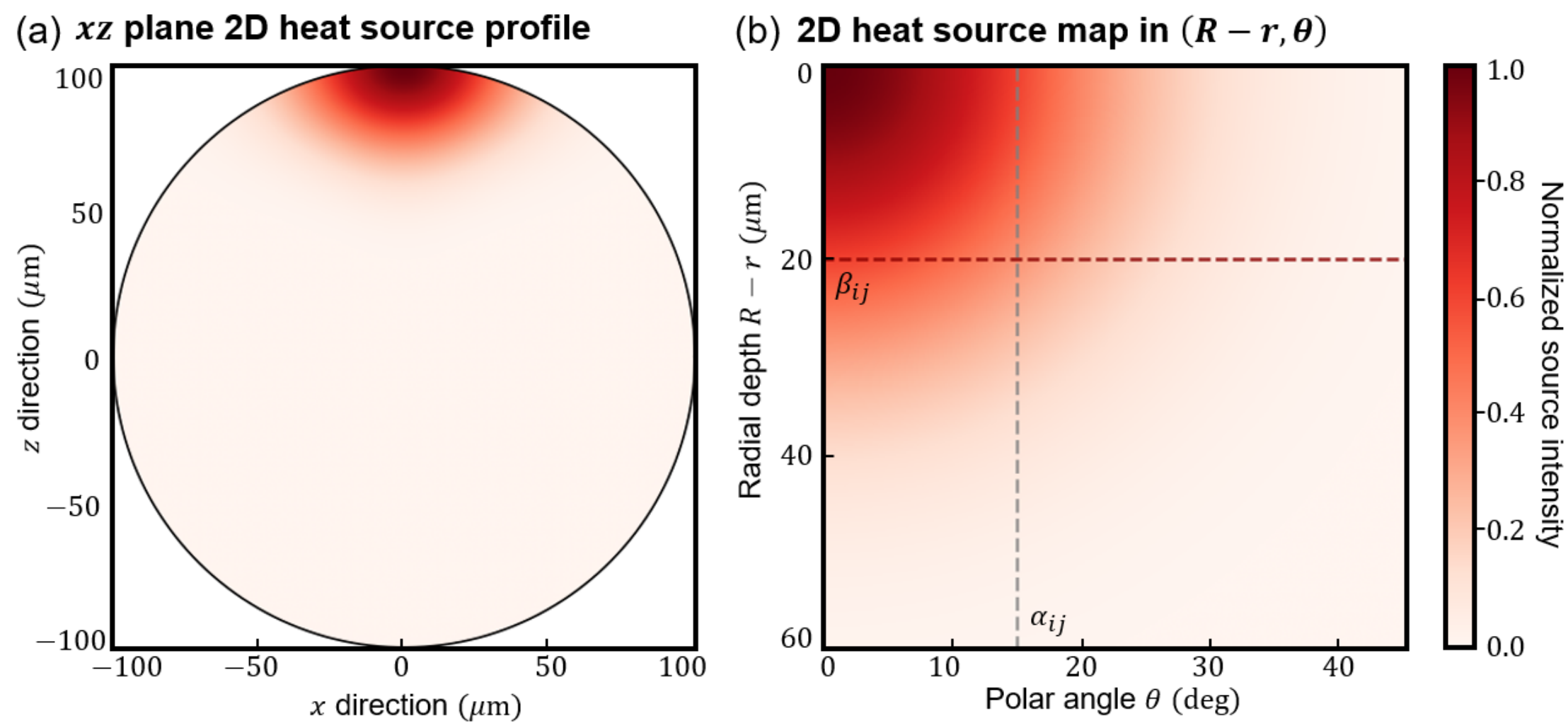

**Figure S2.** Localized Gaussian heat source shape illustration. (a) Two-dimensional heat source distribution in the $xz$ plane for a collision centered at the north pole. (b) The same distribution represented in $(R-r,\theta)$ coordinates. Representative Gaussian widths of $\beta_{ij} = 20$ µm and $\alpha_{ij} = 15°$ are used to illustrate the radial and angular localization, respectively. The color scale denotes the normalized source intensity. The broadened widths are used only for visualization and are not the pair-dependent widths employed in numerical calculations.

### B. Heat partition between two colliding particles

During an inelastic collision, part of the relative kinetic energy of two particles is dissipated through microscopic processes, including atomic vibrations, viscoelastic deformation, and plastic deformation. The loss of translational kinetic energy in a binary normal collision can be expressed in terms of the coefficient of restitution $e$ as

$$\Delta KE_{ij} = \frac{1}{2}\mu_{ij}v_{\mathrm{rel}}^2(1-e^2),$$

where $\mu_{ij} = m_i m_j/(m_i + m_j)$ is the reduced mass and $v_{\mathrm{rel}}$ is the relative normal velocity before collision. The coefficient $e$ characterizes the ratio of relative separation to approach velocities, with $e = 1$ corresponding to an elastic collision and $e < 1$ indicating a loss of translational kinetic energy. Although this relation determines the total kinetic energy loss, it does not uniquely specify how much of the dissipated energy is converted into heat or how that heat is distributed between the two particles.

The actual heat partition depends on the microscopic dissipation processes and the mechanical and material properties of the contacting particles. During impact, the particles undergo local deformation, and the associated mechanical work can be dissipated through different channels, including lattice vibrations, viscoelastic relaxation, and irreversible deformation. Consequently, particles with different elastic properties, hardness, or energy spectra may absorb different fractions of the generated heat. A quantitative determination of this partition would require a microscopic description of the collision dynamics and energy-dissipation processes, which is beyond the scope of the present thermal diffusion model. In the present study, we adopt equal heat deposition into the two contacting particles, $Q_i = Q_j = Q_{ij}$, as a symmetric reference condition only for simplicity. In real collisions, asymmetric energy partition may modify the quantitative temperature difference and the resulting thermoelectric potential. The influence of material-dependent heat partition remains an open question for future microscopic studies.

### C. Spatial eigenmodes under the Robin boundary condition

To determine the resulting temperature profile inside particle $i$, we define the excess temperature relative to the surrounding medium as $T(r,\theta;t) = T_i(r,\theta;t) - T_\infty$. The heat transfer at the particle surface ($r = R_i$) is described by the Robin boundary condition with $T_\infty = 0$

$$k\frac{\partial T}{\partial r}\bigg|_{r=R_i} + \kappa T(r = R_i, \theta; t) = 0,$$

where $k$ is the thermal conductivity and $\kappa$ is the effective interfacial conductance. The corresponding Biot number is $\mathrm{Bi} = \kappa R/k = R/\lambda$. Because the diffusion operator and the boundary condition are linear, the temperature profile can be expanded in a set of spatial eigenfunctions that satisfy the same boundary condition:

$$T(r,\theta;t) = \sum_{n,l} B_{nl}(t)\, \phi_{nl}(r,\theta)\,. \tag{S5}$$

Here, $B_{nl}(t)$ describes the time-dependent amplitude of each thermal mode, while $\phi_{nl}(r,\theta)$ represents its spatial profile. These spatial basis functions are defined as the regular solutions of the Laplacian eigenvalue problem

$$\nabla^2\phi_{nl} = -\left(\frac{\mu_{nl}}{R_i}\right)^2 \phi_{nl}, \tag{S6}$$

subject to regularity at $r = 0$ and the Robin boundary condition $\left(k\frac{\partial\phi_{nl}}{\partial r} + \kappa\phi_{nl}\right)\Big|_{r=R_i} = 0$. For the axisymmetric heat source considered here, separation of variables gives

$$\phi_{nl}(r,\theta) = j_l\left(\mu_{nl}\frac{r}{R_i}\right)P_l(\cos\theta), \tag{S7}$$

where $j_l$ is the spherical Bessel function of the first kind and $P_l$ is the Legendre polynomial of degree $l$. The index $l$ specifies the angular mode, whereas $n$ labels the discrete radial modes. The dimensionless radial eigenvalue $\mu_{nl}$ is determined by applying the Robin boundary condition to the radial basis function at $r = R_i$. This gives

$$\mu_{nl}j_l'(\mu_{nl}) + \mathrm{Bi}\, j_l(\mu_{nl}) = 0, \tag{S8}$$

meaning that the eigenvalues are controlled by the Biot number.

**D. Projection of the localized heat source onto the spatial eigenmodes**

The localized volumetric heat-input rate defined in Sec. S1.A is expanded in the same spatial eigen-basis:

$$\dot{q}(r,\theta) = \frac{Q_{ij}}{\tau_c\Lambda_{ij}}g_r(r)g_\theta(\theta) = \sum_{n,l}\dot{q}_{nl}\phi_{nl}(r,\theta), \tag{S9}$$

where the source-normalization factor $\Lambda_{ij}$ and the radial-angular Gaussian source profiles are defined as $\Lambda_{ij} = 2\pi\int_0^{R_i} g_r(r')r'^2dr'\int_0^{\pi} g_\theta(\theta')\sin\theta'\,d\theta'$ , $g_r(r') = \exp\left(-\frac{(R-r')^2}{2\beta_{ij}^2}\right)$ , and $g_\theta(\theta') = \exp\left(-\frac{\theta'^2}{2\alpha_{ij}^2}\right)$. Here, $\alpha_{ij}$ and $\beta_{ij}$ characterize the radial and angular localization widths of the heat source, respectively. Because the eigen-functions are orthogonal, the coefficient of each $(n,l)$ mode is obtained by projecting the heat source onto $\phi_{nl}$:

$$\dot{q}_{nl} = \frac{\int_V \dot{q}(r,\theta)\phi_{nl}(r,\theta)dV}{\int_V \phi_{nl}^2(r,\theta)dV}. \tag{S10}$$

Under azimuthal symmetry, the volume element is $dV = 2\pi r^2\sin\theta\,drd\theta$, and the numerator becomes

$$\int_V \dot{q}(r,\theta)\phi_{nl}(r,\theta)dV = \frac{2\pi Q_{ij}}{\tau_c\Lambda_{ij}}\int_0^{R_i} g_r(r)j_l\left(\mu_{nl}\frac{r}{R_i}\right)r^2dr\int_0^{\pi} g_\theta(\theta)P_l(\cos\theta)\sin\theta\,d\theta.$$

The denominator can be separated into radial and angular normalization integrals:

$$\int_V \phi_{nl}^2(r,\theta)dV = 2\pi\int_0^{R_i} j_l^2\left(\mu_{nl}\frac{r}{R_i}\right)r^2dr\int_0^{\pi} P_l^2(\cos\theta)\sin\theta\,d\theta.$$

While the angular normalization of the Legendre polynomial is $\int_0^\pi P_l^2(\cos\theta)\sin\theta\,d\theta = \frac{2}{2l+1}$, we denote the corresponding radial normalization integral as

$$J_{nl} = \int_0^{R_i} j_l^2\left(\mu_{nl}\frac{r}{R_i}\right) r^2 dr\,. \tag{S11}$$

Using the spherical Bessel integral identity and the Robin eigenvalue condition, the radial normalization can be expressed as

$$J_{nl} = \frac{R_i^3}{2\mu_{nl}^2} j_l^2(\mu_{nl})[\mu_{nl}^2 + \mathrm{Bi}_i^2 - \mathrm{Bi}_i - l(l+1)], \tag{S12}$$

where $\mathrm{Bi}_i = R_i/\lambda$. Combining Eqs. (S9–S11), the heat source modal coefficient becomes

$$\dot{q}_{nl} = \frac{Q_{ij}\mu_{nl}^2 a_{nl}}{\tau_c R_i^3}, \tag{S13}$$

where $a_{nl}$ is the dimensionless heat source normalization modal coefficient defined as

$$a_{nl} = \frac{(2l+1)\int_0^{R_i} g_r(r) j_l\left(\mu_{nl}\frac{r}{R_i}\right) r^2 dr \int_0^\pi g_\theta(\theta) P_l(\cos\theta)\sin\theta\,d\theta}{\Lambda_{ij} j_l^2(\mu_{nl})[\mu_{nl}^2 + \mathrm{Bi}_i^2 - \mathrm{Bi}_i - l(l+1)]}. \tag{S14}$$

## S2. General time-dependent temperature profile

### A. Continuous heat pulse model through the linear superposition

The time-dependent temperature response can be constructed by regarding the heat supplied during a collision as a continuous superposition of infinitesimal heat pulses. Because the heat-diffusion equation is linear, the response to these pulses can be combined according to Duhamel's principle. Mathematically, the formal solution of Eq. (1) can be expressed using the Green function,

$$T(r,\theta;t) = \int_{V'}\int_0^t G(r,\theta;r',\theta';t-t')\frac{\dot{q}(r',\theta';t')}{\rho c}dt'\,dV'\,. \tag{S15}$$

Here, $G(r,\theta;r',\theta';t-t')$ is the Green function satisfying the Robin boundary condition.

To represent the heat supplied over the finite collision duration, we divide the contact interval into short time intervals $\tau_\delta \sim 10^{-9}$ s [R5]. Because $\dot{q}(r',\theta')$ is the heat-input rate per unit volume, the energy density deposited during one interval is

$$\Delta q(r',\theta') = \dot{q}(r',\theta')\tau_\delta. \tag{S16}$$

Each energy increment can be represented as an instantaneous $\delta$-like pulse applied at $t' = n\tau_\delta$. The temporal heat source during the collision is therefore written as

$$\dot{q}(r',\theta';t') = \sum_{n\tau_\delta \le t'} \Delta q(r',\theta')\delta(t' - n\tau_\delta) = \tau_\delta \sum_{n\tau_\delta \le t'} \dot{q}(r',\theta')\delta(t' - n\tau_\delta)\,. \tag{S17}$$

Eq. (S17) represents the constant heat-input rate defined in Sec. S1.A as a sequence of short heat pulses. Substituting the pulse-train representation in Eq. (S17) into the Green function solution in Eq. (S15), and using the sampling property of the delta function, the time integral reduces to a discrete sum,

$$T(r,\theta;t) = \sum_{n\tau_\delta \le t} \frac{\tau_\delta}{\rho c} \int_{V'} G(r,\theta;r',\theta';t - n\tau_\delta)\dot{q}(r',\theta')dV'\,. \tag{S18}$$

Each term in Eq. (S18) represents the temperature response to a single pulse applied at $t' = n\tau_\delta$. We denote these microscopic responses as

$$T_{\mathrm{micro}}(r,\theta;t - n\tau_\delta) \equiv \frac{\tau_\delta}{\rho c} \int_{V'} G(r,\theta;r',\theta';t - n\tau_\delta)\dot{q}(r',\theta')dV'\,. \tag{S19}$$

The temperature field during the collision is therefore the temporal superposition of the microscopic responses,

$$T(r,\theta;t) = \sum_{n\tau_\delta \le t} T_{\mathrm{micro}}(r,\theta;t - n\tau_\delta)\,. \tag{S20}$$

Here, $T_{\mathrm{micro}}$ represents the temperature response to a single pulse at $t = 0$, carrying the energy density $\dot{q}(r',\theta')\tau_\delta$. In the limit of $\tau_\delta \ll t$, the discrete superposition of the single-pulse responses can be approximated by the continuous integral

$$T(r,\theta;t) \cong \frac{1}{\tau_\delta} \int_0^t T_{\mathrm{micro}}(r,\theta;t')dt'\,, \qquad 0 < t \le \tau_c. \tag{S21}$$

**B. Eigenfunction expansion of a single heat pulse**

A single microscopic heat pulse applied at $t = 0$ is represented by the source term $\dot{q}(r,\theta)\tau_\delta\delta(t)$. Therefore, its subsequent evolution is governed by the homogeneous heat-diffusion equation,

$$\frac{\partial T_{\mathrm{micro}}(r,\theta;t)}{\partial t} = \frac{k}{\rho c}\nabla^2 T_{\mathrm{micro}}(r,\theta;t) + \frac{\dot{q}(r,\theta)\tau_\delta\delta(t)}{\rho c}\,. \tag{S22}$$

Integrating this equation across $t = 0$ gives the temperature jump induced by the pulse,

$$T_{\text{micro}}(r,\theta;0^+) - T_{\text{micro}}(r,\theta;0^-) = \frac{\dot{q}(r,\theta)\tau_\delta}{\rho c}.$$

Because $T_{\text{micro}}$ denotes the temperature increment generated by this pulse, $T_{\text{micro}}(r,\theta;0^-) = 0$, and hence

$$T_{\text{micro}}(r,\theta;0^+) = \frac{\dot{q}(r,\theta)\tau_\delta}{\rho c}. \tag{S23}$$

For $t > 0$, the instantaneous source term vanishes, and the subsequent evolution of the single-pulse response is governed by the homogeneous heat-diffusion equation,

$$\frac{\partial T_{\text{micro}}(r,\theta;t)}{\partial t} = \frac{k}{\rho c}\nabla^2 T_{\text{micro}}(r,\theta;t), \qquad t > 0, \tag{S24}$$

subject to the same Robin boundary condition introduced in Sec. S1.B. We therefore expand the single-pulse response using the corresponding spatial eigenfunctions,

$$T_{\text{micro}}(r,\theta;t) = \sum_{n,l} b_{nl}(t)\phi_{nl}\,(r,\theta)\,. \tag{S25}$$

The heat source obtained in Sec. S1.C is expanded in the same eigenfunction basis as $\dot{q}(r,\theta) = \sum_{n,l}\dot{q}_{nl}\phi_{nl}(r,\theta)$. Evaluating Eq. (S25) at $t = 0^+$ and using the initial condition in Eq. (S23) gives

$$T_{\text{micro}}(r,\theta;0^+) = \sum_{n,l} b_{nl}(0^+)\phi_{nl}\,(r,\theta) = \frac{\tau_\delta}{\rho c}\sum_{n,l}\dot{q}_{nl}\phi_{nl}\,(r,\theta)\,. \tag{S26}$$

Because of the orthogonality of the eigenfunctions, the initial amplitude of each mode is

$$b_{nl}(0^+) = \frac{\tau_\delta \dot{q}_{nl}}{\rho c}. \tag{S27}$$

Substituting Eq. (S25) into the diffusion equation, Eq. (S24), gives with Eq. (S6)

$$\sum_{n,l}\frac{db_{nl}(t)}{dt}\phi_{nl}\,(r,\theta) = -\frac{k}{\rho c}\sum_{n,l}\left(\frac{\mu_{nl}}{R}\right)^2 b_{nl}(t)\phi_{nl}\,(r,\theta)\,. \tag{S28}$$

Using the orthogonality of the eigenfunctions, each mode evolves independently according to

$$\frac{db_{nl}(t)}{dt} = -\frac{k}{\rho c}\left(\frac{\mu_{nl}}{R}\right)^2 b_{nl}(t). \tag{S29}$$

Together with the initial amplitude in Eq. (S27), the solution is

$$b_{nl}(t) = \frac{\tau_\delta \dot{q}_{nl}}{\rho c} \exp\left[-\frac{k\mu_{nl}^2}{\rho c R^2} t\right]. \tag{S30}$$

Therefore, the temperature reponse to a single microscopic heat pulse is

$$T_{\text{micro}}(r, \theta; t) = \frac{\tau_\delta}{\rho c} \sum_{n,l} \dot{q}_{nl} \phi_{nl}\,(r, \theta) \exp\left[-\frac{k\mu_{nl}^2}{\rho c R^2} t\right]. \tag{S31}$$

**C. Time-dependent temperature and the modal coefficient**

Using Eq. (S13), Eq. (S21) and the shape of $T_{\text{micro}}(r, \theta; t)$ gives

$$\begin{aligned} T(r, \theta; t) &= \frac{Q_{ij}}{\tau_c R k} \sum_{n,l} a_{nl} \phi_{nl}\,(r, \theta) \left[1 - \exp\left(-\frac{k\mu_{nl}^2}{\rho c R^2} t\right)\right] \\ &= \frac{Q_{ij}}{\tau_c R k} \sum_{n,l} a_{nl} \phi_{nl}\,(r, \theta) \left[1 - \exp\left(-\frac{t}{\tau_{nl}}\right)\right], \end{aligned} \tag{S32}$$

where $\tau_{nl} = \rho c R^2/(k\mu_{nl}^2)$. Eq. (S32) directly relates the time-dependent temperature field to the spatial projection of the collision-induced heat source.

Furthermore, Eq. (S24) drives the time-dependent surface potential after a single collision event with duration time $\tau_c$. Using the thermocouple relationship under the open-circuit condition [R6], the surface potential of particle $R_i$ which is colliding with $R_j$ can be written as

$$V(R_i, \theta; \tau_c) = -\frac{Q_{ij}}{\tau_c R_i} \frac{S}{k} \sum_{n,l} a_{nl} \phi_{nl}(R, \theta)\ \left[1 - \exp\left(-\frac{\tau_c}{\tau_{nl}}\right)\right]. \tag{S33}$$

## D. Numerical convergence test of spherical-Bessel eigenmodes and application

For the numerical evaluation of Eq. (S33), the eigenmode summation was truncated at finite angular and radial cutoffs, $n_{\max}$ and $l_{\max}$, respectively. Convergence was tested by increasing both cutoffs in increments of 20. The cutoffs were accepted when the relative changes in the surface-temperature profile, contact temperature, and independently refined angular and radial contributions were all below $10^{-3}$ for two successive refinements. The angular expansion showed only weak variation among the investigated cases, and $l_{\max} \cong 120$ was sufficient to satisfy the prescribed convergence criteria. In contrast, the radial expansion exhibited a pronounced dependence on particle size and material properties, requiring substantially different $n_{\max}$ values among the individual cases (Fig. S3). This difference can be understood from the spatial localization of the collision-induced heat. The Biot number governs the degree of thermal exchange at the particle boundary and thus the extent to which the deposited heat is confined near the particle surface, whereas the thermal diffusivity, $k/\rho c$, determines how rapidly the heat spreads into the particle interior. Under conditions where the collision-induced heat remains more strongly localized near the surface, the radial temperature profile contains sharper spatial variations. Resolving such localized thermal structures with the spherical-Bessel expansion requires a larger number of radial eigenmodes to maintain the same numerical tolerance.

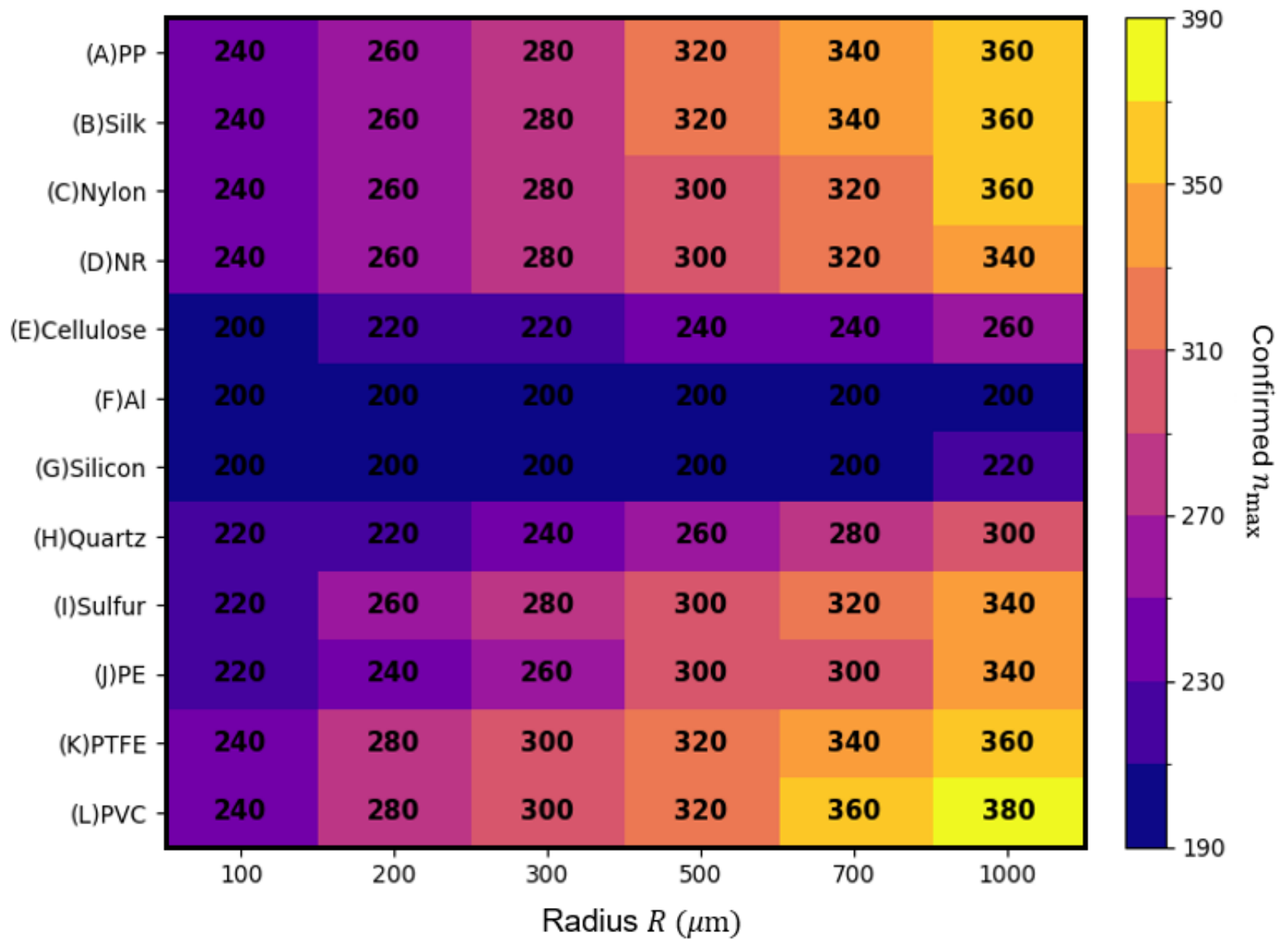

**Figure S3. Radial-basis convergence for different system scales and materials.** The particle radius $R$ denotes the radius of an equal-sized collision pair ($R_i = R_j = R$). Colors and labels indicate the confirmed $n_{\max}$ satisfying the convergence criterion defined in Sec. S2.D. For a fixed collision duration $\tau_c$, increasing $R$ reduces the relative diffusion depth $\delta_{\text{th}}/R$, where $\delta_{\text{th}} \sim \sqrt{\gamma \tau_c}$, and therefore requires more radial modes. Material dependence arises from differences in thermal diffusivity $\gamma = k/\rho c$.

Accordingly, the required $n_{\max}$ was determined separately for each material-radius combination through the convergence test, rather than applying a single uniform radial cutoff to all cases.

### E. Numerical verification of the boundary condition limits

As a final consistency check, we examine whether the Robin solution continuously recovers the Neumann and Dirichlet limits. The limiting forms of the Robin eigenvalue condition are immediate: $\mathrm{Bi} \to 0$ and $\mathrm{Bi} \to \infty$ recover the Neumann and Dirichlet eigenvalue equations, respectively. Convergence of the eigenvalues alone, however, does not guarantee convergence of the reconstructed finite-time temperature field because the eigenfunctions and source-projection coefficients also depend on the boundary condition. In addition, the lowest Robin eigenvalue vanishes in the Neumann limit, requiring explicit verification that its modal contribution remains finite and recovers the uniform zero mode.

We distinguish the three boundary conditions by the superscript $b \in \{\mathrm{N}, \mathrm{R}, \mathrm{D}\}$, where N=Neumann, R=Robin, and D=Dirichlet. Their radial eigenvalues at the particle surface satisfy

$$\begin{gathered} \mu_{nl}^{(\mathrm{R})} j_l'\left(\mu_{nl}^{(\mathrm{R})}\right) + \mathrm{Bi}\, j_l\left(\mu_{nl}^{(\mathrm{R})}\right) = 0, \\ j_l'\left(\mu_{nl}^{(\mathrm{N})}\right) = 0, \qquad j_l\left(\mu_{nl}^{(\mathrm{D})}\right) = 0. \end{gathered} \tag{S34}$$

The corresponding eigenfunctions are defined as in Eq. (S7),

$$\phi_{nl}^{(b)}(r,\theta) = j_l\left(\mu_{nl}^{(b)} \frac{r}{R_i}\right) P_l(\cos\theta),$$

with the radial normalization as in Eq. (S11)

$$J_{nl}^{(b)} = \int_0^{R_i} j_l^2\left(\mu_{nl}^{(b)} \frac{r}{R_i}\right) r^2 dr.$$

The same localized heat-input rate is projected independently onto each boundary-dependent eigen basis. Following Eq. (S10), the resulting modal coefficient is

$$\begin{aligned} \dot{q}_{nl}^{(b)} &= \frac{Q_{ij}}{\tau_c \Lambda_{ij}} \frac{2l+1}{2J_{nl}^{(b)}} \times \int_0^{R_i} g_r(r) j_l\left(\mu_{nl}^{(b)} \frac{r}{R_i}\right) r^2 dr \\ &\quad \times \int_0^{\pi} g_\theta(\theta) P_l(\cos\theta) \sin\theta \, d\theta \,, \end{aligned} \tag{S35}$$

where $g_r$ and $g_\theta$, and $\Lambda_{ij}$ are the source profile and normalization factor defined in Eq. (S9). Thus, the physical heat source is identical in all three calculations, whereas its modal representation changes according to the boundary conditions.

The Neumann solution is then

$$T^{(\mathrm{N})}(r,\theta;t) = \frac{3Q_{ij}}{4\pi\rho c R^3}\frac{t}{\tau_c} + \sum_{(n,l)\neq(0,0)} \frac{\dot{q}_{nl}^{(\mathrm{N})} R_i^2}{k\left(\mu_{nl}^{(\mathrm{N})}\right)^2} \phi_{nl}^{(\mathrm{N})}(r,\theta)\left[1-\exp\left(-\frac{k\left(\mu_{nl}^{(\mathrm{N})}\right)^2}{\rho c R_i^2}t\right)\right], \quad \text{(S36)}$$

where the first term is the spatially uniform Neumann zero mode. Its coefficient follows from

$$\dot{q}_{00}^{(\mathrm{N})} = \frac{\int_V \dot{q}(r,\theta)dV}{\int_V dV} = \frac{3Q_{ij}}{4\pi\tau_c R^3}. \quad \text{(S37)}$$

For a general mode, Eq. (S29) including the continuous heat-input term can be written as

$$\frac{dB_{nl}^{(b)}}{dt} = -\frac{k}{\rho c}\left(\frac{\mu_{nl}^{(b)}}{R_i}\right)^2 B_{nl}^{(b)} + \frac{\dot{q}_{nl}^{(b)}}{\rho c}. \quad \text{(S38)}$$

For the Neumann zero mode, $\mu_{00}^{(\mathrm{N})} = 0$, so the diffusion-relaxation term vanishes:

$$\frac{dB_{00}^{(\mathrm{N})}}{dt} = \frac{\dot{q}_{00}^{(\mathrm{N})}}{\rho c}. \quad \text{(S39)}$$

Using $B_{00}^{(\mathrm{N})}(0) = 0$ and $\phi_{00}^{(\mathrm{N})}(0) = 1$, the zero-mode temperature contribution is consequently

$$T_{00}^{(\mathrm{N})}(r,\theta;t) = B_{00}^{(\mathrm{N})}(t)\phi_{00}^{(\mathrm{N})}(0) = \frac{3Q_{ij}}{4\pi\rho c R^3}\frac{t}{\tau_c}, \qquad 0 < t \leq \tau_c. \quad \text{(S40)}$$

Similarly, we can construct the Robin and Dirichlet solution as

$$T^{(\mathrm{R})}(r,\theta;t) = \sum_{n,l} \frac{\dot{q}_{nl}^{(\mathrm{R})} R_i^2}{k\left(\mu_{nl}^{(\mathrm{R})}\right)^2} \phi_{nl}^{(\mathrm{R})}(r,\theta)\left[1-\exp\left(-\frac{k\left(\mu_{nl}^{(\mathrm{R})}\right)^2}{\rho c R_i^2}t\right)\right], \quad \text{(S41a)}$$

$$T^{(\mathrm{D})}(r,\theta;t) = \sum_{n,l} \frac{\dot{q}_{nl}^{(\mathrm{D})} R_i^2}{k\left(\mu_{nl}^{(\mathrm{D})}\right)^2} \phi_{nl}^{(\mathrm{D})}(r,\theta)\left[1-\exp\left(-\frac{k\left(\mu_{nl}^{(\mathrm{D})}\right)^2}{\rho c R_i^2}t\right)\right]. \quad \text{(S41b)}$$

Using Eq. (S36) and (S41), we verify the limiting consistency of the Robin solution over a broad range of Biot numbers. The lowest $l = 0$ Robin-mode contribution at the particle surface, normalized by the uniform Neumann zero mode, approaches unity as Bi → 0 and vanishes as Bi → ∞ [Fig. S4(a)]. For the complete temperature field, the volume-weighted RMS deviation $\varepsilon^{(b)} = \left[V^{-1}\int_V \left|T^{(\mathrm{R})} - T^{(b)}\right|^2 dV\right]^{1/2} / \left[V^{-1}\int_V \left|T^{(b)}\right|^2 dV\right]^{1/2}$ ($b \in \{\mathrm{N}, \mathrm{D}\}$), decreases toward zero in the corresponding Neumann and Dirichlet limits [Fig. S4(b)]. These results confirm both the regular recovery of the Neumann zero mode and the global convergence of the finite-time Robin solution to the two limiting boundary conditions.

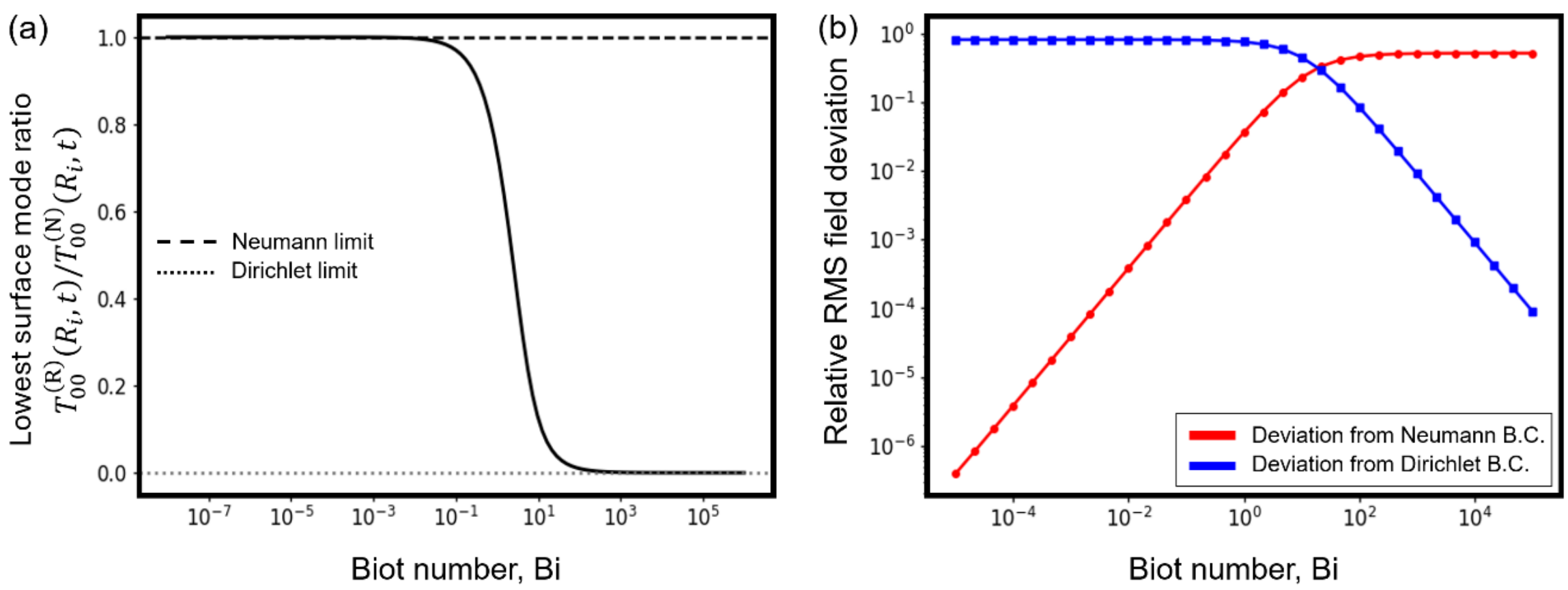

**Figure S4. Numerical verification of the Neumann and Dirichlet limits of the Robin solution.** (a) Surface contribution of the lowest $l = 0$ Robin mode, $T_{00}^{(\mathrm{R})}(R_i; t)$, normalized by the uniform Neumann zero-mode contribution, $T_{00}^{(\mathrm{N})}(R_i; t) = \frac{3Q_{ij}}{4\pi\rho c R^3}\frac{t}{\tau_c}$, as a function of the Biot number. The ratio approaches unity in the Neumann limit and vanishes in the Dirichlet limit. (b) Relative volume-weighted RMS error of the complete finite-time Robin solution with respect to the Neumann and Dirichlet reference fields. The reference fields are evaluated directly from their respective eigenvalue problems Eq. (S34). The errors decrease toward their respective limits as $\mathrm{Bi} \to 0$ and $\mathrm{Bi} \to \infty$. Calculations are performed for a dimensionless unit sphere with a normalized Gaussian source of $\alpha = 10°$ and $\beta/R_i = 0.2$ at $t/\tau_D = 10^{-2}$, and using $(n_{\max}, l_{\max}) = (20,10)$.

## S3. Reduction of the multi-collision response to a heat-capacity model

### A. Statistical averaging of randomly-oriented collisions and lowest-mode approximation

In repeated random collisions, the total thermal field is formed by the superposition of many such localized pulses with different contact orientations. Because these orientations are statistically random, the directional contributions of the $l \neq 0$ modes do not build up coherently and are suppressed in the orientationally averaged response. The mean angular response is therefore determined by the spherically symmetric $l = 0$ component. To evaluate the contribution of the remaining radial modes to the particle-averaged temperature, we consider the volume average of the eigenfunction defined in Eq. (S7):

$$\langle \phi_{n0} \rangle_V = \frac{1}{V} \int_V j_0 \left( \mu_{n0} \frac{r}{R_i} \right) P_0(\cos\theta) dV = \frac{3}{R_i^3} \int_0^{R_i} j_0 \left( \mu_{n0} \frac{r}{R_i} \right) r^2 dr$$

$$= \frac{3(\sin\mu_{n0} - \mu_{n0}\cos\mu_{n0})}{\mu_{n0}^3}. \tag{S42}$$

Using the Robin eigenvalue condition in Eq. (S8), this becomes

$$\langle \phi_{n0} \rangle_V = \frac{3\mathrm{Bi}_i \sin\mu_{n0}}{\mu_{n0}^3}. \tag{S43}$$

For the lowest mode, the Robin eigenvalue approaches $\mu_{00}^2 \simeq 3\mathrm{Bi}_i$ as $\mathrm{Bi}_i \to 0$, yielding $\langle \phi_{00} \rangle_V \to 1$. In contrast, the higher radial eigenvalues remain finite, and their volume-averaged contributions approach zero as the Bi decreases: $\langle \phi_{n0} \rangle_V \to 0$ $(n \geq 1)$. Thus, in the Neumann-like Robin limit ($\mathrm{Bi}_i \ll 1$), the particle-averaged response can be approximated by the lowest mode ($n = l = 0$), even when nonuniform higher modes have not completely relaxed.

Applying this lowest-mode approximation to Eq. (2) of the main text, the particle-averaged potential associated with a single collision between particle $i$ and $j$ becomes

$$V_i^{(ij)}(\tau_c) \simeq -\frac{SQ_{ij}}{kR_i\tau_c} a_{00} \langle \phi_{00} \rangle_V \left( 1 - \exp\left[ -\frac{\tau_c}{\tau_{00}} \right] \right). \tag{S44}$$

For $\tau_c \ll \tau_{00}$, the first-order expansion $1 - e^{-\tau_c/\tau_{00}} \simeq \tau_c/\tau_{00}$ and the relax time $\tau_{00} = \rho c R^2/(k\mu_{00}^2)$ gives Eq. (3) of the main text. The remaining spatial contribution can be evaluated directly from the heat-source projection coefficient $a_{nl}$ derived in Eq. (S14). For $n = l = 0$, this coefficient reduces

$$a_{00} = \frac{I_r(\mu_{00}) I_\theta}{\Lambda_{ij} j_0^2(\mu_{00})[\mu_{00}^2 + \mathrm{Bi}_i^2 - \mathrm{Bi}_i]}, \tag{S45}$$

where, using the radial and angular source profile already defined in Sec. S1, $I_r(\mu_{00}) = \int_0^{R_i} g_r(r) j_0\left(\mu_{00}\frac{r}{R_i}\right) r^2 dr$ and $I_\theta = \int_0^\pi g_\theta(\theta)\sin\theta\, d\theta$. In the weak heat-loss limit, $\mu_{00}^2 \simeq 3\mathrm{Bi}$, $j_0(\mu_{00}) \simeq 1$, and therefore $\mu_{00}^2 + \mathrm{Bi}_i^2 - \mathrm{Bi}_i \simeq \frac{2}{3}\mu_{00}^2$. Consequently,

$$\lim_{\mathrm{Bi}_i \to 0} \mu_{00}^2\, a_{00} = \frac{3}{2}\frac{I_r(0) I_\theta}{\Lambda_{ij}}.$$

Using the heat-source normalization introduced in Sec. S1, $\Lambda_{ij} = 2\pi I_r(0) I_\theta$, we obtain

$$\lim_{\mathrm{Bi}_i \to 0} \mu_{00}^2\, a_{00} \langle \phi_{00} \rangle_V = \frac{3}{4\pi}. \tag{S46}$$

Substituting this limit into Eq. (3) of the main text yields the single-collision heat-capacity response, which provides the basis for the statistical accumulation model in Eq. (4).

For the multi-collision calculation, we prescribe the effective thermal contribution per collision as $Q_{ij}^{(\mathrm{mc})} = Q_0^{(\mathrm{mc})} A_{ij}$, with $Q_0^{(\mathrm{mc})} = 10^{-5}\ \mathrm{J\ cm^{-2}}$. The normalized collision probability is therefore $P_{ij} = P(R_i, R_j) = \frac{(R_i + R_j)^2}{\sum_{i<j}(R_i+R_j)^2}$. For an expected total number of collision events $N_{\mathrm{tot}}$, the number of collisions between particle $i$ and $j$ is independently sampled from $N_{ij} \sim \mathrm{Poisson}(N_{\mathrm{tot}} P_{ij})$. The accumulated temperature rise of particle $i$, $\Delta T_i$, is therefore

$$\Delta T_i = \frac{1}{C_i}\sum_{j\neq i} N_{ij}\, Q_{ij}^{(\mathrm{mc})} = \sum_{j \neq i} T_i^{(ij)}.$$

Because the thermal response is linear, the accumulated temperature of particle $i$ can equivalently be decomposed into contributions from individual collision pairs $(i, j)$. The corresponding pair-specific potential contribution is

$$V_i^{(ij)} = -S T_i^{(ij)} = -\frac{S N_{ij}\, Q_{ij}^{(\mathrm{mc})}}{C_i}, \tag{S47}$$

which gives Eq. (4) of the main text upon substituting the particle heat capacity.

### B. Charge transfer and surface charge density calculation

To estimate the resulting charge transfer, we introduce an effective charge-transfer capacitance $C_{\mathrm{tr}} = 0.1\ \mathrm{\mu C V^{-1}}$. We use this coefficient as a size-independent scaling parameter that relates the induced voltage to the accumulated charge. Accordingly, the charge transfer for a collision pair $(i, j)$ is given by

$$q_{ij} = C_{\mathrm{tr}}\left[V_i^{(ij)} - V_j^{(ij)}\right] = -\frac{3C_{\mathrm{tr}}N_{ij}Q_{ij}^{(\mathrm{mc})}}{4\pi}\frac{S}{\rho c}\left(\frac{1}{R_i^3} - \frac{1}{R_j^3}\right). \tag{S48}$$

Charge conservation is imposed for every collision pair by assigning opposite charge increments to the two particles: $\Delta q_i^{(ij)} = -q_{ij}$, $\Delta q_j^{(ij)} = q_{ij}$. The final accumulated charge of particle $i$ is obtained by summing up the signed contributions from all its collision partners: $q_i = \sum_{j \neq i} \Delta q_i^{(ij)}$.

The corresponding surface charge density is

$$\sigma_i = \frac{q_i}{4\pi R_i^2}. \tag{S49}$$

The accumulated charge is reported as the surface charge density, consistent with the experimentally observed surface-area scaling of stationary granular charge distributions [R7]. For quartz, the negative Seebeck coefficient produces the conventional polarity trend in which smaller particles become negatively charged and larger particles positively charged. For nylon, the positive Seebeck coefficient reverses the polarity, causing smaller particles to become positively charged and larger particles negatively charged. Thus, the model describes size-dependent charge segregation through the combined effects of pairwise collision statistics, contact geometry, particle heat capacity, and the sign of Seebeck coefficient. The statistical robustness of these results was further examined by varying the random seed and the total number of collision events [Fig. S5]. The group-averaged charge separation was insensitive to the sampled realization, while the expected polarity fraction approached a stable value with increasing collision count, supporting the use of $N_{\mathrm{tot}} = 10^4$ in Fig. 4 of the main text.

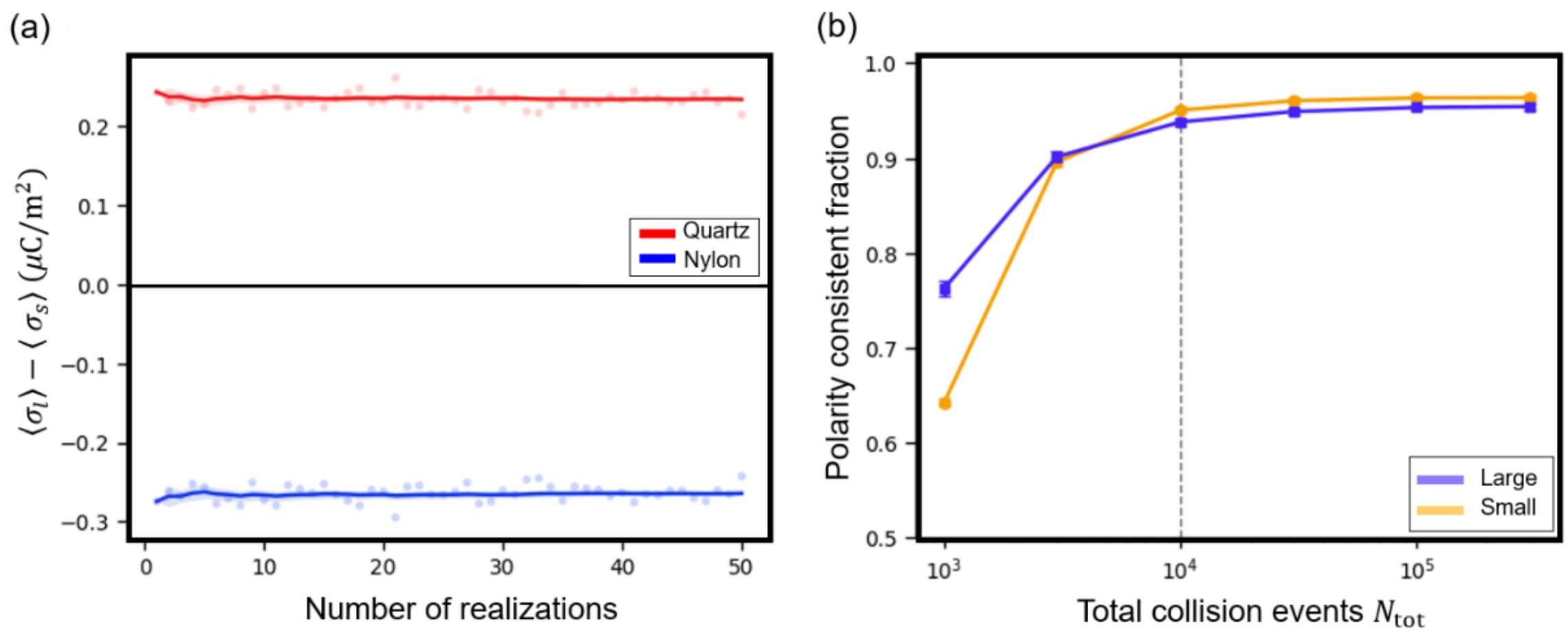

**Figure S5. Statistical robustness of the multi-collision charge-segregation model.** (a) Convergence of the group-averaged surface charge density separation, $\langle \sigma_l \rangle - \langle \sigma_s \rangle$. Pale points show individual realizations; solid lines show the cumulative meaning. The opposite signs for nylon and quartz arise from their opposite Seebeck sign. (b) Fraction of small and large particles acquiring the material-dependent expected polarity as a function of the total number of collision event $N_{\mathrm{tot}}$. The symbols and error bars represent the mean and 95% confidence interval over 30 independent realizations, respectively. The vertical dashed line marks $N_{\mathrm{tot}} = 10^4$, the condition used in Fig. 4 of the main text. The polarity-consistent fractions increase with collision number and approach stable values, confirming that the predicted charge segregation trend is robust against both random sampling and the selected collision total count.

**References**

[R1] V. L. Popov, M. Heß, and E. Willert, *Handbook of Contact Mechanics* (Springer, Berlin, 2018).

[R2] V. L. Popov, *Contact Mechanics and Friction: Physical Principles and Applications* (Springer, Berlin, 2010).

[R3] D. L. Blair and A. Kudrolli, *Collision statistics of driven granular materials*, Phys. Rev. E **67**, 041301 (2003).

[R4] J. R. Royer, D. J. Evans, L. Oyarte, Q. Guo, E. Kapit, M. E. Möbius, S. R. Waitukaitis, and H. M. Jaeger, *High-speed tracking of rupture and clustering in freely falling granular streams*, Nature **459**, 1110 (2009).

[R5] J.-H. Mun, E.-C. Shin, J. Seo, and Y.-H. Kim, *Triboelectric charge transfer theory driven by thermoelectric effect*, Phys. Rev. Research **7**, L032067 (2025).

[R6] E.-C. Shin, J.-H. Ko, H.-K. Lyeo, and Y.-H. Kim, *Derivation of a governing rule in triboelectric charging and series from thermoelectricity*, Phys. Rev. Research **4**, 023131 (2022).

[R7] P. S. Iliev, F. K. Wittel, and H. J. Herrmann, *Evolution of fragment size distributions from the crushing of granular materials*, Phys. Rev. E **99**, 012904 (2019).